%% file: main.tex
\pdfoutput=1

\documentclass[manuscript,screen,table]{acmart}

\AtBeginDocument{%
  \providecommand\BibTeX{{%
    \normalfont B\kern-0.5em{\scshape i\kern-0.25em b}\kern-0.8em\TeX}}}

\usepackage{booktabs}
\usepackage{multirow}
\usepackage{array}
\usepackage{mathtools}
\usepackage{longtable}
\usepackage{wasysym}
\usepackage{enumitem}

\usepackage{tikz}
\usepackage{lipsum}
\usepackage{subfig}
\usepackage{graphicx}

\newcommand*\rot{\rotatebox{90}}

\copyrightyear{2021}
\acmYear{2021}
\setcopyright{acmlicensed}\acmConference[CHI '21]{CHI Conference on Human Factors in Computing Systems}{May 8--13, 2021}{Yokohama, Japan}
\acmBooktitle{CHI Conference on Human Factors in Computing Systems (CHI '21), May 8--13, 2021, Yokohama, Japan}
\acmPrice{15.00}
\acmDOI{10.1145/3411764.3445610}
\acmISBN{978-1-4503-8096-6/21/05}

\begin{document}



\title{What Makes a Dark Pattern... Dark?} 
\subtitle{Design Attributes, Normative Considerations, and Measurement Methods}

\author{Arunesh Mathur}
\email{amathur@princeton.edu}
\affiliation{Princeton University}

\author{Jonathan Mayer}
\email{jonathan.mayer@princeton.edu}
\affiliation{Princeton University}

\author{Mihir Kshirsagar}
\email{mihir@princeton.edu}
\affiliation{Princeton University}

\renewcommand{\shortauthors}{Mathur et al.}


\begin{abstract}
  \input{abstract.tex}
\end{abstract}

\maketitle

\section{Introduction}
\label{sec:introduction}
\input{sections/introduction.tex}


\section{A Comparison of Dark Pattern Definitions, Types, and Attributes}
\label{sec:background}
\input{sections/background.tex}

\section{Choice Architecture in Behavioral Economics, Philosophy, and Law}
\label{sec:disciplines}

\input{sections/disciplines.tex}

\section{Normative Perspectives on Dark Patterns}
\label{sec:norm}
\input{sections/normative.tex}

\section{Measurement Methods and Applications}
\label{sec:hmeasure}
\input{sections/measure.tex}



\section{Conclusion}
\label{sec:conclusion}
\input{sections/conclusion.tex}

\begin{acks}
We thank our shepherd, Aneesha Singh, for her thoughtful guidance. A previous version of this paper appeared at the 2020 Privacy Law Scholars Conference (PLSC). We thank the participants at the workshop for their comments and feedback. We also thank Arvind Narayanan, Marshini Chetty, Ari Ezra Waldman, and Elena Lucherini for their advice and feedback.
\end{acks}

\bibliographystyle{ACM-Reference-Format}
\bibliography{bibliography}



\end{document}

%% file: abstract.tex
There is a rapidly growing literature on dark patterns, user interface designs---typically related to shopping or privacy---that researchers deem problematic. Recent work has been predominantly descriptive, documenting and categorizing objectionable user interfaces. These contributions have been invaluable in highlighting specific designs for researchers and policymakers. But the current literature lacks a conceptual foundation: What makes a user interface a dark pattern? \textit{Why} are certain designs problematic for users or society?

We review recent work on dark patterns and demonstrate that the literature does not reflect a singular concern or consistent definition, but rather, a set of thematically related considerations. Drawing from scholarship in psychology, economics, ethics, philosophy, and law, we articulate a set of normative perspectives for analyzing dark patterns and their effects on individuals and society. We then show how future research on dark patterns can go beyond subjective criticism of user interface designs and apply empirical methods grounded in normative perspectives.

%% file: sections/introduction.tex
Recent scholarship has called attention to dark patterns, user interface designs that researchers deem problematic. The preponderance of academic literature on dark patterns has curated collections of objectionable user interface designs~\cite{brignull, gray-dark-patterns-2018} and highlighted the frequency of dark patterns in specific contexts, such as privacy settings~\cite{bosch-tales-2016}, online gaming~\cite{zagal2013dark}, and online shopping~\cite{mathur2019cscw}. Related work has also traced the lineage of dark patterns to discrete trends in business and social science~\cite{narayanan2020dark}.

Meanwhile, legislators and regulators are taking notice. In the U.S., Congress is considering legislation that would restrict dark patterns~\cite{detour-act}, and California~\cite{caballot} recently enacted legislation (via a ballot initiative) that defines and prohibits dark patterns associated with privacy consent. In the EU, regulators are beginning to target dark patterns for enforcement actions~\cite{frobrukerradet-deceived-2018,cnil}. Dark patterns are also a persistent source of public frustration, such as on Twitter (\texttt{\#darkpattern} and \texttt{@darkpatterns}) and a Reddit subreddit (\texttt{r/assholedesign}) with over 1.5 million members.

The current academic discourse about dark patterns is built on a shaky foundation, however. While the literature has rich descriptive contributions, calling attention to user interface designs that the authors (understandably) find objectionable, the contours and normative underpinnings of dark patterns have received less attention. What, exactly, makes a user interface design a dark pattern? And what makes a dark pattern normatively problematic?

This gap in the literature has significant consequences. Within the Human--Computer Interaction (HCI) community and related areas of computer science, publications are in limited dialogue, and there is a disconnect between research results and possible government responses. Meanwhile, outside of computer science, critics have begun to dismiss dark patterns research as academic handwringing about aggressive marketing and sharp dealing---the online equivalents of practices that are unsavory but commonplace for offline businesses~\cite{cric1, cric2}. Absent a clearly articulated rationale or standard for taking action against certain user interface designs, legislators and regulators on both sides of the Atlantic will continue to struggle with possible market interventions~\cite{ukreport}. 


This paper aims to provide analytical clarity and normative foundations for dark patterns research in HCI, other areas of computer science, and related disciplines. We begin with a review of the rapidly growing literature on dark patterns, identifying key definitions and taxonomic concepts. We demonstrate that dark patterns research is not grounded in a singular, unified concern or definition. Rather, dark patterns scholarship---similar to privacy scholarship---is motivated by a family of thematically related considerations. We illustrate this point by synthesizing dark patterns definitions, types, and taxonomies from recent scholarship into a pair of themes.

Next, we discuss how topics in related disciplines can enrich the dark patterns discourse in the HCI community, including sludges in behavioral economics, manipulation in philosophy and ethics, and market manipulation in law. We draw out the perspectives and normative considerations that these other disciplines offer, and we use these to develop a set of normative lenses for evaluating dark patterns.

Finally, we describe how future research can use HCI  methods to examine dark patterns through each normative lens. We demonstrate this path forward for future work by applying our recommended approach in a case study. We conclude with a discussion of how the normative perspectives and measurement methods we develop in this paper can be of use to HCI practitioners and policymakers.

Our primary contributions are as follows.

\begin{itemize}
\item We review the dark patterns literature in HCI and related areas of computer science, identify diverse definitions and taxonomies, and synthesize cross-cutting themes (Section~\ref{sec:background}). 
\item We connect the themes in dark patterns scholarship to representative related work in other disciplines (Section~\ref{sec:disciplines}). 
\item We describe a set of discrete normative perspectives, grounded in the representative related work, that could guide future dark patterns research (Section~\ref{sec:norm}).
\item We propose that future dark patterns research use methods that align with those normative perspectives, and we demonstrate our proposed approach with a case study (Section~\ref{sec:hmeasure}).
\end{itemize}

%% file: sections/background.tex
We begin with a scoping review of the dark patterns academic literature, which is predominantly (but not exclusively) in the area of HCI.\footnote{We refer to this analysis as a scoping review rather than a systematization of the literature, because our goal is to evaluate key definitions and concepts in publications rather than reported evidence.} To facilitate this survey, we compiled a dataset of papers by searching the ACM Digital Library, arXiv, and Google Scholar for work that referenced ``dark patterns'' or similar terms.\footnote{The complete set of search queries was: ``anti-pattern(s),'' ``dark pattern(s),'' ``deceptive design pattern(s),'' ``FoMo design(s),'' and ``manipulative design pattern(s).'' We selected these queries based on the terminology in dark patterns scholarship we had already identified, and we believe they captured a representative---if not fully complete---dataset of related work.} We then filtered the dataset, retaining papers that a) discuss dark patterns in the context of user interface design, and b) have been published in an academic venue.\footnote{We did not include papers that only incidentally discussed dark patterns. We included the initial unpublished Brignull work~\cite{brignull}, a concurrent publication by Conti and Sobiesk~\cite{conti2010malicious} that predates the popularization of the ``dark patterns'' term, and a forthcoming paper by Luguri and Strahilevitz~\cite{luguri2019shining} because those contributions have generated significant discussion.} The resulting dataset included a total of 20 papers, published in HCI conferences and journals (e.g., CHI, NordiCHI, and DIS), security and privacy venues (e.g., PETS and NSPW), law journals, and a psychology journal.\footnote{The dataset is available at \url{https://github.com/citp/dark-patterns}. We compiled the dataset in December 2020.}

Next, we augmented our dataset with legislation and regulatory materials related to dark patterns.\footnote{We included introduced legislation and official agency publications in our dataset. Our search additionally encompassed press releases, hearing materials, and statements from individual members of multi-member commissions; we reviewed this content to ensure we had not missed materials for the dataset.} We ran a search for ``dark pattern(s)'' on WestLaw for federal and state materials in the U.S., and we identified international materials from references in publications. Our search yielded the Deceptive Experiences to Online Users Reduction (DETOUR) Act~\cite{detour-act}, the California Privacy Rights Act (CPRA, also known as Proposition 24 in the November 2020 election)~\cite{caballot}, a report by France's data protection agency (CNIL)~\cite{cnil}, and a report by the Norwegian Consumer Council (NCC/Frobrukerr\r{a}det)~\cite{frobrukerradet-deceived-2018}.

We then coded each document in the dataset for four distinct properties. First, we examined dark pattern definitions. We coded a document as introducing a definition if it included a definition, explicitly or implicitly, and the definition was either entirely original to the work or went beyond a verbatim reference to prior work. If a document introduced a definition, we extracted the definition. The second property we coded was whether a document offered a typology or taxonomy of dark patterns. If it did, we extracted the categories of dark patterns, individual dark patterns, and associated descriptions. The third property that we coded was whether a document discussed any normative considerations, either explicitly or implicitly. If it did, we extracted the normative concepts. Finally, we coded each document's use of user studies. If a paper included a new user study, we extracted the goals of the study, the type of study, and any measurement metrics.

In the following subsections, we use our dataset to compare dark patterns definitions (Section~\ref{sec:definitionsdarkpatterns}), types  (Section~\ref{sec:dark_pattern_types}), and attributes  (Section~\ref{sec:dark_pattern_attributes}) in prior work. We trace the origins and evolution of various dark patterns definitions, types, and attributes. We show that the literature---which we emphasize is valuable---has a number of inconsistencies and contradictions. We argue that these are a result of the literature's attempt to capture dark patterns in one succinct definition or list of types. Instead, we derive a pair of themes related to choice architecture that succinctly reflect the dark patterns discourse (Section~\ref{sec:dark_patterns_themes}): modifying the user's choice set and manipulating the information flow to the user.

\subsection{Defining Dark Patterns}
\label{sec:definitionsdarkpatterns}

\def\arraystretch{1.4}
\setlength{\tabcolsep}{7pt}
\begin{table}[t]
\centering
\caption{A classification of various ``dark pattern'' definitions in academic literature, law, and policy. Documents are ordered by date.}
\label{tab:def}
\resizebox{\textwidth}{!}{
\begin{tabular}{@{}rrcccccccccccccccccccc@{}}
\toprule

& & \multicolumn{15}{c}{\textbf{Academic Publications}} & & \multicolumn{4}{c}{\textbf{\begin{tabular}[c]{@{}c@{}}Government\\ Materials\end{tabular}}}\\
\cmidrule(lr){3-17}
\cmidrule(l){19-22}
                                                                  &    & \rot{\textbf{Brignull~\cite{brignull}}} &    \rot{\textbf{Conti \& Sobiesk~\cite{conti2010malicious}}} & \rot{\textbf{Zagal et al. ~\cite{zagal2013dark}}} & \rot{\textbf{Lewis~\cite{lewis-irresistible-apps-2014}}} & \rot{\textbf{B{\"o}sch et al.~\cite{bosch-tales-2016}}} & \rot{\textbf{Gray et al.~\cite{gray-dark-patterns-2018}}} & \rot{\textbf{Mathur et al.~\cite{mathur2019cscw}}} &
                                                \rot{\textbf{Luguri \& Strahilevitz~\cite{luguri2019shining}}}   & 
                                                \rot{\textbf{Lacey \& Caudwell~\cite{cuteness}}}   &

                                    \rot{\textbf{Utz et al.~\cite{utz2019informed}}} & 
                                    \rot{\textbf{Westin \& Chiasson~\cite{fomo}}} &
                                    \rot{\textbf{Waldman~\cite{WALDMAN2020105}}} &
                                    \rot{\textbf{Day \& Stemler~\cite{darkanticompetitive}}} &
                                                \rot{\textbf{Gray et al.~\cite{gray-asshole}}} &
                                                \rot{\textbf{Maier \& Harr~\cite{maier2020dark}}} & &\rot{\textbf{NCC~\cite{frobrukerradet-deceived-2018}}} & \rot{\textbf{CNIL~\cite{cnil}}} & \rot{\textbf{DETOUR Act~\cite{detour-act}}} & \rot{\textbf{CPRA~\cite{caballot}}} \\ \cmidrule(lr){2-17}
\cmidrule(l){19-22}

\multirow{8}{*}{Characteristics of the User Interface} & Coercive & & & & & & & \RIGHTcircle & & & & & & & \RIGHTcircle & & & & & & \\

  & \cellcolor[HTML]{F9F9F9}Deceptive & \cellcolor[HTML]{F9F9F9} & \cellcolor[HTML]{F9F9F9} & \cellcolor[HTML]{F9F9F9} & \cellcolor[HTML]{F9F9F9} & \cellcolor[HTML]{F9F9F9} & \cellcolor[HTML]{F9F9F9}\CIRCLE & \cellcolor[HTML]{F9F9F9}\RIGHTcircle & \cellcolor[HTML]{F9F9F9} & \cellcolor[HTML]{F9F9F9} & \cellcolor[HTML]{F9F9F9} & \cellcolor[HTML]{F9F9F9} & \cellcolor[HTML]{F9F9F9} & \cellcolor[HTML]{F9F9F9} & \cellcolor[HTML]{F9F9F9}\RIGHTcircle & \cellcolor[HTML]{F9F9F9} & \cellcolor[HTML]{F9F9F9} & \cellcolor[HTML]{F9F9F9} &  \cellcolor[HTML]{F9F9F9} & \cellcolor[HTML]{F9F9F9} & \cellcolor[HTML]{F9F9F9} \\

  & Malicious & & \CIRCLE & & & & & & & & & & & & & & & & & & \\
  
  & \cellcolor[HTML]{F9F9F9}Misleading & \cellcolor[HTML]{F9F9F9} & \cellcolor[HTML]{F9F9F9} & \cellcolor[HTML]{F9F9F9} & \cellcolor[HTML]{F9F9F9} & \cellcolor[HTML]{F9F9F9}\CIRCLE & \cellcolor[HTML]{F9F9F9} & \cellcolor[HTML]{F9F9F9} & \cellcolor[HTML]{F9F9F9} & \cellcolor[HTML]{F9F9F9} & \cellcolor[HTML]{F9F9F9} & \cellcolor[HTML]{F9F9F9} & \cellcolor[HTML]{F9F9F9} & \cellcolor[HTML]{F9F9F9} & \cellcolor[HTML]{F9F9F9} & \cellcolor[HTML]{F9F9F9} & \cellcolor[HTML]{F9F9F9} & \cellcolor[HTML]{F9F9F9} & \cellcolor[HTML]{F9F9F9}\CIRCLE & \cellcolor[HTML]{F9F9F9} & \cellcolor[HTML]{F9F9F9}\\
  
  &  Obnoxious &  &  &  &  &  &  &  &  &  &  &  &  &  & \RIGHTcircle &  &  &  &  &  &  \\
  
  & \cellcolor[HTML]{F9F9F9}Seductive & \cellcolor[HTML]{F9F9F9} & \cellcolor[HTML]{F9F9F9} & \cellcolor[HTML]{F9F9F9} & \cellcolor[HTML]{F9F9F9} & \cellcolor[HTML]{F9F9F9} & \cellcolor[HTML]{F9F9F9} & \cellcolor[HTML]{F9F9F9} & \cellcolor[HTML]{F9F9F9} & \cellcolor[HTML]{F9F9F9} & \cellcolor[HTML]{F9F9F9} & \cellcolor[HTML]{F9F9F9} & \cellcolor[HTML]{F9F9F9} & \cellcolor[HTML]{F9F9F9} & \cellcolor[HTML]{F9F9F9} & \cellcolor[HTML]{F9F9F9}\CIRCLE & \cellcolor[HTML]{F9F9F9} & \cellcolor[HTML]{F9F9F9} & \cellcolor[HTML]{F9F9F9} & \cellcolor[HTML]{F9F9F9} & \cellcolor[HTML]{F9F9F9}\\
  
  & Steering &  &  &  &  &  &  & \RIGHTcircle &  &  &  &  &  &  &  &  &  & \CIRCLE &  &  &  \\
  
  & \cellcolor[HTML]{F9F9F9}Trickery & \cellcolor[HTML]{F9F9F9}\CIRCLE & \cellcolor[HTML]{F9F9F9} & \cellcolor[HTML]{F9F9F9} & \cellcolor[HTML]{F9F9F9} & \cellcolor[HTML]{F9F9F9} & \cellcolor[HTML]{F9F9F9} & \cellcolor[HTML]{F9F9F9} & \cellcolor[HTML]{F9F9F9} & \cellcolor[HTML]{F9F9F9} & \cellcolor[HTML]{F9F9F9} & \cellcolor[HTML]{F9F9F9} & \cellcolor[HTML]{F9F9F9}\CIRCLE & \cellcolor[HTML]{F9F9F9} & \cellcolor[HTML]{F9F9F9} & \cellcolor[HTML]{F9F9F9} & \cellcolor[HTML]{F9F9F9} & \cellcolor[HTML]{F9F9F9} & \cellcolor[HTML]{F9F9F9} & \cellcolor[HTML]{F9F9F9} & \cellcolor[HTML]{F9F9F9} \\
  
  \cmidrule(lr){2-17}
\cmidrule(l){19-22}

\multirow{13}{*}{Mechanisms of Effect on Users} &  Attack users &  & \RIGHTcircle &  &  &  &  &  &  &  &  &  &  &  &  &  &  &  &  &  &  \\

& \cellcolor[HTML]{F9F9F9}Confuse users & \cellcolor[HTML]{F9F9F9} & \cellcolor[HTML]{F9F9F9} & \cellcolor[HTML]{F9F9F9} & \cellcolor[HTML]{F9F9F9} & \cellcolor[HTML]{F9F9F9} & \cellcolor[HTML]{F9F9F9} & \cellcolor[HTML]{F9F9F9} & \cellcolor[HTML]{F9F9F9}\RIGHTcircle & \cellcolor[HTML]{F9F9F9} & \cellcolor[HTML]{F9F9F9} & \cellcolor[HTML]{F9F9F9} & \cellcolor[HTML]{F9F9F9} & \cellcolor[HTML]{F9F9F9} & \cellcolor[HTML]{F9F9F9} & \cellcolor[HTML]{F9F9F9} & \cellcolor[HTML]{F9F9F9} & \cellcolor[HTML]{F9F9F9} & \cellcolor[HTML]{F9F9F9} & \cellcolor[HTML]{F9F9F9} & \cellcolor[HTML]{F9F9F9} \\

& Deceive users &  &  &  &  &  &  &  &  & \CIRCLE &  &  &  &  &  &  &  &  &  &  &  \\

& \cellcolor[HTML]{F9F9F9}Exploit users & \cellcolor[HTML]{F9F9F9}  & \cellcolor[HTML]{F9F9F9}\RIGHTcircle & \cellcolor[HTML]{F9F9F9} & \cellcolor[HTML]{F9F9F9} & \cellcolor[HTML]{F9F9F9} & \cellcolor[HTML]{F9F9F9} & \cellcolor[HTML]{F9F9F9} & \cellcolor[HTML]{F9F9F9} & \cellcolor[HTML]{F9F9F9} & \cellcolor[HTML]{F9F9F9} & \cellcolor[HTML]{F9F9F9} & \cellcolor[HTML]{F9F9F9} & \cellcolor[HTML]{F9F9F9} & \cellcolor[HTML]{F9F9F9} & \cellcolor[HTML]{F9F9F9} & \cellcolor[HTML]{F9F9F9} & \cellcolor[HTML]{F9F9F9} & \cellcolor[HTML]{F9F9F9} & \cellcolor[HTML]{F9F9F9} & \cellcolor[HTML]{F9F9F9} \\
 
& Manipulate users &      & \RIGHTcircle &   &  &  &  &  & \RIGHTcircle &  &  & \CIRCLE & \CIRCLE &  &  &  &  &  &  &  &  \\

& \cellcolor[HTML]{F9F9F9}Mislead users & \cellcolor[HTML]{F9F9F9} & \cellcolor[HTML]{F9F9F9} & \cellcolor[HTML]{F9F9F9} & \cellcolor[HTML]{F9F9F9} & \cellcolor[HTML]{F9F9F9} & \cellcolor[HTML]{F9F9F9} & \cellcolor[HTML]{F9F9F9} & \cellcolor[HTML]{F9F9F9} & \cellcolor[HTML]{F9F9F9} & \cellcolor[HTML]{F9F9F9} & \cellcolor[HTML]{F9F9F9} & \cellcolor[HTML]{F9F9F9} & \cellcolor[HTML]{F9F9F9} & \cellcolor[HTML]{F9F9F9} & \cellcolor[HTML]{F9F9F9}\CIRCLE & \cellcolor[HTML]{F9F9F9} & \cellcolor[HTML]{F9F9F9} & \cellcolor[HTML]{F9F9F9} & \cellcolor[HTML]{F9F9F9} & \cellcolor[HTML]{F9F9F9} \\

& Steer users &  &  &  &  &  &  &  &  &  &  &  &  & \CIRCLE &  &  &  &  &  &  & \\

& \cellcolor[HTML]{F9F9F9}Subvert user intent & \cellcolor[HTML]{F9F9F9}\CIRCLE & \cellcolor[HTML]{F9F9F9} & \cellcolor[HTML]{F9F9F9} & \cellcolor[HTML]{F9F9F9}\CIRCLE & \cellcolor[HTML]{F9F9F9}\CIRCLE & \cellcolor[HTML]{F9F9F9} & \cellcolor[HTML]{F9F9F9} & \cellcolor[HTML]{F9F9F9} & \cellcolor[HTML]{F9F9F9} & \cellcolor[HTML]{F9F9F9} & \cellcolor[HTML]{F9F9F9}\CIRCLE & \cellcolor[HTML]{F9F9F9}\CIRCLE & \cellcolor[HTML]{F9F9F9}\CIRCLE & \cellcolor[HTML]{F9F9F9} & \cellcolor[HTML]{F9F9F9} & \cellcolor[HTML]{F9F9F9} & \cellcolor[HTML]{F9F9F9} & \cellcolor[HTML]{F9F9F9} & \cellcolor[HTML]{F9F9F9} & \cellcolor[HTML]{F9F9F9} \\

& Subvert user preferences &  &  &  &  & \CIRCLE &  & \CIRCLE & \RIGHTcircle &  &  &  &  &  &  &  &  &  & \CIRCLE & \CIRCLE & \CIRCLE \\

& \cellcolor[HTML]{F9F9F9}Trick users & \cellcolor[HTML]{F9F9F9} & \cellcolor[HTML]{F9F9F9} & \cellcolor[HTML]{F9F9F9} & \cellcolor[HTML]{F9F9F9} & \cellcolor[HTML]{F9F9F9}\CIRCLE & \cellcolor[HTML]{F9F9F9} & \cellcolor[HTML]{F9F9F9} & \cellcolor[HTML]{F9F9F9} & \cellcolor[HTML]{F9F9F9} & \cellcolor[HTML]{F9F9F9} & \cellcolor[HTML]{F9F9F9}\CIRCLE & \cellcolor[HTML]{F9F9F9} & \cellcolor[HTML]{F9F9F9} & \cellcolor[HTML]{F9F9F9} & \cellcolor[HTML]{F9F9F9}\CIRCLE & \cellcolor[HTML]{F9F9F9} & \cellcolor[HTML]{F9F9F9} & \cellcolor[HTML]{F9F9F9} & \cellcolor[HTML]{F9F9F9} & \cellcolor[HTML]{F9F9F9}\\

& Undermine user autonomy &  &  &  &  &  &  &  &  &  &  &  &  &  &  &  &  &  &  & \CIRCLE & \CIRCLE \\

& \cellcolor[HTML]{F9F9F9}Without user consent & \cellcolor[HTML]{F9F9F9} & \cellcolor[HTML]{F9F9F9} & \cellcolor[HTML]{F9F9F9}\CIRCLE & \cellcolor[HTML]{F9F9F9} & \cellcolor[HTML]{F9F9F9} & \cellcolor[HTML]{F9F9F9} & \cellcolor[HTML]{F9F9F9} & \cellcolor[HTML]{F9F9F9} & \cellcolor[HTML]{F9F9F9} & \cellcolor[HTML]{F9F9F9} & \cellcolor[HTML]{F9F9F9} & \cellcolor[HTML]{F9F9F9} & \cellcolor[HTML]{F9F9F9} & \cellcolor[HTML]{F9F9F9} & \cellcolor[HTML]{F9F9F9} & \cellcolor[HTML]{F9F9F9} & \cellcolor[HTML]{F9F9F9} & \cellcolor[HTML]{F9F9F9} & \cellcolor[HTML]{F9F9F9} & \cellcolor[HTML]{F9F9F9}\\

& Without user knowledge &  &  &  &  &  &  &  &  &  &  &  &  &  &  &  &  &  & \CIRCLE &  &  \\

  \cmidrule(lr){2-17}
\cmidrule(l){19-22}

 \multirow{2}{*}{Role of User Interface Designers} & \cellcolor[HTML]{F9F9F9}Abuse of designer knowledge & \cellcolor[HTML]{F9F9F9} & \cellcolor[HTML]{F9F9F9} & \cellcolor[HTML]{F9F9F9} & \cellcolor[HTML]{F9F9F9} & \cellcolor[HTML]{F9F9F9} & \cellcolor[HTML]{F9F9F9}\CIRCLE & \cellcolor[HTML]{F9F9F9} & \cellcolor[HTML]{F9F9F9} & \cellcolor[HTML]{F9F9F9}\CIRCLE & \cellcolor[HTML]{F9F9F9} & \cellcolor[HTML]{F9F9F9}\CIRCLE & \cellcolor[HTML]{F9F9F9} & \cellcolor[HTML]{F9F9F9} & \cellcolor[HTML]{F9F9F9} & \cellcolor[HTML]{F9F9F9}\CIRCLE & \cellcolor[HTML]{F9F9F9} & \cellcolor[HTML]{F9F9F9} & \cellcolor[HTML]{F9F9F9} & \cellcolor[HTML]{F9F9F9} & \cellcolor[HTML]{F9F9F9}\\
 
&  Designer intent &  & \CIRCLE & \CIRCLE &  &  & \CIRCLE &   & \CIRCLE &  &  &  &  &  & \CIRCLE &  &  &  & \CIRCLE & \CIRCLE & \CIRCLE \\
 
  \cmidrule(lr){2-17}
\cmidrule(l){19-22}

 \multirow{2}{*}{Benefits and Harms} & \cellcolor[HTML]{F9F9F9}Benefit to service & \cellcolor[HTML]{F9F9F9} & \cellcolor[HTML]{F9F9F9}\CIRCLE & \cellcolor[HTML]{F9F9F9} & \cellcolor[HTML]{F9F9F9} & \cellcolor[HTML]{F9F9F9} & \cellcolor[HTML]{F9F9F9} & \cellcolor[HTML]{F9F9F9}\CIRCLE & \cellcolor[HTML]{F9F9F9} & \cellcolor[HTML]{F9F9F9} & \cellcolor[HTML]{F9F9F9}\CIRCLE & \cellcolor[HTML]{F9F9F9} & \cellcolor[HTML]{F9F9F9} & \cellcolor[HTML]{F9F9F9} & \cellcolor[HTML]{F9F9F9}\CIRCLE & \cellcolor[HTML]{F9F9F9} & \cellcolor[HTML]{F9F9F9} & \cellcolor[HTML]{F9F9F9} & \cellcolor[HTML]{F9F9F9} & \cellcolor[HTML]{F9F9F9} & \cellcolor[HTML]{F9F9F9} \\
 
& Harm to users &  &  & \CIRCLE &  &  & \CIRCLE &  &  &  &  &  & \CIRCLE &  &  &  &  & \CIRCLE &  &  & \\
 \bottomrule
\end{tabular}}
\par
\bigskip
\mbox{}\hfill \CIRCLE \hspace{5pt}Required element of ``dark pattern'' definition \hspace{10pt} \RIGHTcircle \hspace{5pt}Alternative element of ``dark pattern'' definition

\end{table}

In 2010, when Brignull first introduced the term ``dark patterns'' on the website \texttt{darkpatterns.org}~\cite{brignull}, he described the user interface designs as ``tricks used in websites and apps that make you do things that you didn't mean to, like buying or signing up for something.'' Brignull's initial work set a flurry of academic research into motion that attempted to define and describe dark patterns. We identified a total of 19 such definitions in our dataset, and from these definitions, we inductively surfaced four facets of dark pattern definitions.

The first facet describes characteristics of the user interface that can affect users. Some definitions describe dark pattern user interfaces as design ``tricks'' (Brignull~\cite{brignull} and Waldman~\cite{WALDMAN2020105}), while others describe them as ``misleading'' interfaces (B{\"o}sch et al.~\cite{bosch-tales-2016} and CNIL~\cite{cnil}). Some documents define dark patterns with a set of alternative characteristics. For instance, Mathur et al.~\cite{mathur2019cscw} state that dark pattern designs are ``coercing, steering, or deceiving,'' and Gray et al.~\cite{gray-asshole} note that dark patterns have ``obnoxious, coercive, or deceitful'' properties.

The second facet of definitions is the mechanism of effect for influencing users.\footnote{We recognize that the distinction between characteristics of the user interface and mechanisms of effect on users may, in some instances, be more grammatical than substantive. We parse this distinction because some readers may find it meaningful.} Collectively, the dark pattern definitions specify 13 distinct ways that user interfaces could influence users. Some definitions describe dark patterns as subverting user intent (Brignull~\cite{brignull}, Lewis~\cite{lewis-irresistible-apps-2014}, B{\"o}sch et al.~\cite{bosch-tales-2016}, Westin and Chiasson~\cite{fomo}, Waldman~\cite{WALDMAN2020105}, and Day and Stemler~\cite{darkanticompetitive}) or subverting user preferences (B{\"o}sch et al.~\cite{bosch-tales-2016}, Mathur et al.~\cite{mathur2019cscw}, Luguri and Strahilevitz~\cite{luguri2019shining}, CNIL~\cite{cnil}, DETOUR Act~\cite{detour-act}, and CPRA~\cite{caballot}). Some documents state that dark patterns ``trick'' users (B{\"o}sch et al.~\cite{bosch-tales-2016}, Westin and Chiasson~\cite{fomo}, and Maier and Harr~\cite{maier2020dark}), while others define dark patterns as undermining user autonomy (DETOUR Act~\cite{detour-act} and CPRA~\cite{caballot}). Many definitions specify multiple mechanisms of effect on users. For instance, Westin and Chiasson~\cite{fomo} define dark patterns to ``manipulate'' users, subvert user intent, and ``trick'' users, whereas Conti and Sobiesk~\cite{conti2010malicious} state that dark patterns ``attack,'' ``exploit,'' or ``manipulate'' users.

The third facet of dark pattern definitions is the role of the user interface designer. Some definitions involve a designer abusing their domain-specific knowledge of human behavior (Gray et al.~\cite{gray-dark-patterns-2018}, Lacey and Caudwell~\cite{cuteness}, Westin and Chiasson~\cite{fomo}, and Maier and Harr~\cite{maier2020dark}). Other definitions state that designers intentionally deploy dark patterns to achieve a goal (Conti and Sobiesk~\cite{conti2010malicious}, Zagal et al.~\cite{zagal2013dark}, Gray et al.~\cite{gray-dark-patterns-2018}, Luguri and Strahilevitz~\cite{luguri2019shining}, Gray et al.~\cite{gray-asshole}, CNIL~\cite{cnil}, DETOUR Act~\cite{detour-act}, and CPRA~\cite{caballot}). We note that, as far back as 1998, Fogg articulated a very similar concept: user interface designers could use ``persuasive technologies'' to intentionally influence users~\cite{fogg1999persuasive}.

The fourth facet of dark pattern definitions is the benefits and harms resulting from a user interface design. Some definitions describe a dark pattern as aiming to benefit an online service (Conti and Sobiesk~\cite{conti2010malicious}, Mathur et al.~\cite{mathur2019cscw}, Utz et al.~\cite{utz2019informed}, and Gray et al.~\cite{gray-asshole}). Other dark pattern definitions involve harm to users (Zagal et al.~\cite{zagal2013dark}, Gray et al.~\cite{gray-dark-patterns-2018}, Waldman~\cite{WALDMAN2020105}, and the NCC~\cite{frobrukerradet-deceived-2018}).

We summarize the 19 dark pattern definitions along these four facets in Table~\ref{tab:def}. As the table shows, there is significant variation among the facets reflected in definitions. For instance, nine definitions do not involve any characteristic of the user interface, four definitions do not specify a mechanism of effect on users, eight definitions do not address the role of user interface designers, and ten definitions do not involve benefit or harm elements.

There is also significant variation within each facet of the definitions. For example, Brignull~\cite{brignull} and Conti and Sobiesk~\cite{conti2010malicious} both have definitions that involve characteristics of user interfaces, but the former describes dark patterns as ``tricks'' while the latter notes that dark patterns are ``malicious.''

Another point of divergence is that, even when definitions share an element, the element may be required for one definition but not another definition. Both Gray et al.~\cite{gray-dark-patterns-2018} and Mathur et al.~\cite{mathur2019cscw}, for example, take the position that deception is an element of a dark pattern. But deception is a required element for the former, while one of several alternative elements for the latter. Similarly, B{\"o}sch et al.~\cite{bosch-tales-2016} describes dark patterns as subverting user preferences, as does Luguri and Strahilevitz~\cite{luguri2019shining}. But the latter work additionally considers user interfaces that confuse or manipulate users to be dark patterns.

A final challenge for dark patterns definitions is lack of specificity in recurring terminology. Brignull~\cite{brignull} and Waldman~\cite{WALDMAN2020105}, for example, describe dark patterns as involving ``tricks.'' But what constitutes a trick? Similarly, Maier and Harr~\cite{maier2020dark} note that dark patterns are ``seductive'' user interfaces. What makes a user interface seductive?

\subsection{Types of Dark Patterns}
\label{sec:dark_pattern_types}

\begin{figure}[t]
    \centering
    \setlength{\fboxsep}{1pt}%
\setlength{\fboxrule}{0.4pt}%
\fbox{
    \includegraphics[width=0.5\textwidth]{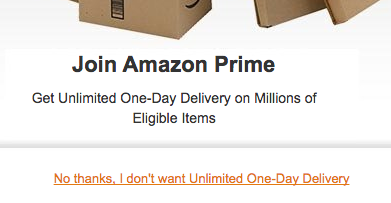}}
    \caption{The Confirmshaming dark pattern on Amazon. The option to decline the offer uses framing to guilt the user into opting in.}
    \label{fig:dp}
    \Description{The Confirmshaming dark pattern on Amazon. The option to decline the offer uses framing to guilt the user into opting in.}
\end{figure}

Prior academic work on dark patterns has often gone beyond a definition and provided a taxonomy of user interface types, with a rich description of each dark pattern. Brignull's~\cite{brignull} initial taxonomy introduced 12 types of dark patterns. Bait-and-Switch, for example, is a user interface where ``the user sets out to do one thing, but a different, undesirable thing happens instead.'' Confirmshaming is ``the act of guilting the user into opting in to something.'' Figure~\ref{fig:dp} illustrates one instance of the Confirmshaming dark pattern. In concurrent work, Conti and Sobiesk~\cite{conti2010malicious} documented 11 categories of problematic user interfaces. Obfuscation, for instance, involves ``hiding desired information and interface elements.'' Trick involves ``misleading the user or other attempts at deception, such as spoofed content or interface elements.''

Several subsequent projects have focused on dark patterns in specific categories of applications or systems. Zagal et al.~\cite{zagal2013dark} described seven types of dark patterns used by video game developers. The authors introduced dark patterns such as Pay to Skip, where a developer charges users to solve in-game challenges and creates an uneven playing field for paying and non-paying users. Grinding requires users to perform repetitive and tedious tasks in a game, which can keep users invested in playing. Lewis~\cite{lewis-irresistible-apps-2014} focused on dark patterns in mobile apps. Temporal dark patterns in apps ``request too much time, provide too little entertainment time, or result in users feeling like they've wasted their time.'' Monetary dark patterns cause ``users to regret spending money, lose track of how much money they will spend, or [not] know how much money will be required in order to progress.'' Social Capital dark patterns cause ``users [to] stand the risk of losing their social standing, or feel like they're interacting with software out of social obligation to fulfill a motivation.'' Greenberg et al.~\cite{greenberg-proxemics-2014} considered systems that account for physical space and described a taxonomy of eight ``proxemic'' (i.e., physical proximity) interaction dark patterns. The Attention Grabber, for example, is where a user ``happens to pass by the field of view of a strategically located system, and the system takes deliberate action to attract and keep that person's attention.'' The Milk Factor is where a system ``forces the user to move through or go to a specific location in order to get a service.'' Finally, Lacey and Caudwell~\cite{cuteness} focused on home robotics systems and whether ``cuteness'' can constitute a dark pattern.

Another set of studies has documented dark patterns that undermine user privacy. B{\"o}sch et al.~\cite{bosch-tales-2016} introduced a set of seven privacy dark patterns, including Bad Defaults, where ``the default options are sometimes chosen badly in the sense that they ease or encourage the sharing of personal information.'' Hidden Legalese Stipulations is when ``services hide stipulations in the policies which target the privacy of the user.'' The NCC report~\cite{frobrukerradet-deceived-2018} on dark patterns described a set of five privacy dark patterns. Ease, for example, is ``making the process toward [privacy-respecting] alternatives a long and arduous process.'' Framing is when a user interface ``focus[es] on the positive aspects of one choice, while glossing over any potentially negative aspects, inclin[ing] many users to comply with the service provider's wishes.'' The CNIL report~\cite{cnil} emphasized how dark patterns can undermine data protection rights, documenting 18 distinct types of dark patterns. Impenetrable Wall, for instance, is a user interface that ``blocks access to a service [with] a cookie wall or account creation [when that] is not necessary to use the service.'' Last Minute Consent involves ``seeking consent for the collection of data at a specific moment where we know that the individual is in a weak position because [they are] in a hurry or impatient to finish.''

Finally, Gray et al.~\cite{gray-dark-patterns-2018} created a taxonomy of dark patterns from the ground up by collecting a corpus of dark patterns submitted by users on Twitter. The authors revamped Brignull's original taxonomy of dark patterns and introduced new dark patterns. Nagging, for instance, is when ``the user's desired task is interrupted one or more times by other tasks not directly related to the one the user is focusing on.'' Forced Action is when ``users are required to perform a specific action to access (or continue to access) specific functionality.'' In another study, Gray et al.~\cite{gray-asshole} created a similar taxonomy by collecting and analyzing a corpus of user interface designs submitted by users on Reddit. The authors introduced Misrepresenting, for example, in which ``designers provide ambiguous and incorrect information in a direct way to trick users.'' Controlling is when ``designers interrupt or restrict the user’s task ﬂow, explicitly directing the task flow.''

Recent studies have also attempted to measure the prevalence of dark patterns---those that appeared in the taxonomies above---on the web and in mobile apps. Using automated web crawling methods, Mathur et al.~\cite{mathur2019cscw} identified 1,800 dark patterns on shopping websites including dark patterns such as Sneaking, Social Proof, and Countdown Timers. Utz et al.~\cite{utz2019informed} analyzed the cookie banner notices of a random sample of 1,000 of the most popular websites in the EU, finding that over 50\% of the banners contained at least one dark pattern. The dark patterns that the authors identified included privacy-unfriendly defaults, hidden opt-out choices, and preselected checkboxes that enable data collection. In another measurement study focused on Android apps, Di Geronimo et al.~\cite{dig} found that more than 95\% of the 200 most popular apps contain at least one dark pattern from the Gray et al~\cite{gray-dark-patterns-2018} taxonomy. Nouwens et al.~\cite{nouwens} examined 680 of the most popular consent management platforms---services that help websites manage user privacy consent---on the top 10,000 websites in the U.K., finding that only 11.8\% of websites do not contain any dark patterns. Soe et al.~\cite{circumvention} manually collected and analyzed 300 data collection consent notices from Scandinavian and English news websites. The authors found that nearly every notice (297) contained at least one dark pattern.

Combining the diverse range of dark pattern taxonomies and types with the diverse range of definitions sharply illustrates the lack of conceptual consistency in prior work. The B{\"o}sch et al.~\cite{bosch-tales-2016} and CNIL~\cite{cnil} definitions, for example, provide that a dark pattern must be misleading. But there is nothing inherently misleading about many of the dark patterns described in the literature, such as Conti and Sobiesk's~\cite{conti2010malicious} Coercion and Forced Work patterns, Gray et al.'s~\cite{gray-dark-patterns-2018} Forced Action dark pattern, or Mathur et al.'s~\cite{mathur2019cscw} Obstruction pattern. These user interface designs can be entirely truthful with users.

What's more, there is some conceptual inconsistency even within papers---projects can describe dark pattern types that do not appear to fit within the provided definition. Brignull~\cite{brignull} and B{\"o}sch et al.~\cite{bosch-tales-2016}, for example, describe dark patterns as ``tricks.'' It is difficult to see where the trick is, however, in Brignull's Confirmshaming or B{\"o}sch et al.'s Forced Registration---these user interface designs are often entirely (and frustratingly) transparent to users. Similarly, Zagal et al.~\cite{zagal2013dark} describe dark patterns as user interface designs that likely lack consent. The Grinding dark pattern discussed in the paper, however, inherently relies on the user's continuing consent in playing the game. Gray et al.~\cite{gray-dark-patterns-2018} define dark patterns as ``deceptive functionality.'' But there is nothing inherently deceptive about the paper's Nagging dark pattern.

In sum, the dark pattern definitions and taxonomies in prior work have been exceedingly valuable for surfacing descriptive insights and calling attention to problematic practices. But the dark patterns literature currently lacks a clear and consistent conceptual foundation.

\subsection{Attributes of Dark Patterns}
\label{sec:dark_pattern_attributes}

A recent paper by Mathur et al.~\cite{mathur2019cscw} attempted to take a step beyond dark pattern definitions and taxonomies, offering a set of shared higher-level attributes that could descriptively organize instances of dark patterns in the literature. We discuss the Mathur et al. attributes below, and we propose an additional attribute.

\subsubsection{Asymmetric} Asymmetric dark patterns impose unequal burdens on the choices available to the user. The choices that benefit the service are feature prominently while the options that benefit the user are typically tucked away behind several clicks or are obscured from view by varying the style and position of the choice. Asymmetric dark patterns are particularly common in consent interfaces. For instance, the Trick Questions dark pattern can impose a cognitive burden on choices that withhold consent by using confusing language and double negatives. The Confirmshaming dark pattern can use emotion to burden choices, associating guilt with certain choices and not others. Privacy Zuckering burdens choices with user interface friction, hiding privacy-respecting settings behind obscure menus.


\subsubsection{Covert} Covert dark patterns push a user toward selecting certain decisions or outcomes, but hide the influence mechanism from the user. Some dark patterns achieve this property by exploiting cognitive biases, while others use color and style to steer users. Dark patterns that fit within Brignull's~\cite{brignull} Misdirection category are often covert. For instance, a shopping website might add a ``free'' gift to the user's shopping cart (e.g., a magazine subscription) to inflate the apparent shopping cart discount in a way that is not readily apparent. Websites may leverage the decoy effect cognitive bias, in which an additional choice---the decoy---makes certain other choices seem more appealing. Users may fail to recognize the decoy's presence is merely to influence their decision making.

\subsubsection{Deceptive} Deceptive dark patterns induce false beliefs in users through affirmative misstatements, misleading statements, or omissions. Deceptive dark patterns are particularly common on shopping websites. For instance, the Countdown Timer dark pattern involves a phony deadline for a deal or offer, with a prominent ticking timer showing that the deadline is imminent. Mathur et al.~\cite{mathur2019cscw} observed that over 40\% of all the timers they discovered on shopping websites either reset on expiration or were inconsequential to the advertised offer.

\subsubsection{Information hiding} Information hiding dark patterns obscure or delay the presentation of necessary information to users. One example is the Sneaking category of dark patterns. Another is the Hidden Subscription dark pattern, which does not reveal to users that the purchase they are making is a recurring subscription. Similarly, the Hidden Costs dark pattern hides relevant cost information until right before the user completes the transaction and is unlikely to change their course of action.

\subsubsection{Restrictive} Restrictive dark patterns reduce or eliminate the choices presented to users. For example, the Forced Action dark pattern on a website might require users to agree to both the terms of use as well as receiving marketing emails before creating an account. Similarly, the Hard to Cancel dark pattern makes it easy for users to sign up for subscription services online, but does not give the user an online option to cancel that same service.
\def\arraystretch{1.2}
\begin{table*}[t]
\centering
\caption{Dark patterns attributes grouped by how they modify the user's choice architecture.}
\label{tab:dp}
\resizebox{\textwidth}{!}{
\begin{tabular}{@{}llp{0.5\textwidth}@{}}
\toprule
\multicolumn{1}{l}{\textbf{Choice Architecture}}   & \multicolumn{1}{l}{\textbf{Attribute}} & \multicolumn{1}{l}{\textbf{Description}}                                                                    \\\midrule

Modifying the decision space                       & \cellcolor[HTML]{F9F9F9}Asymmetric                             & \cellcolor[HTML]{F9F9F9}Unequal burdens on choices available to the user                                      \\
                                                   & Restrictive                            & Eliminate certain choices that should be available to users                                                 \\
                                                   & \cellcolor[HTML]{F9F9F9}Disparate Treatment                                & \cellcolor[HTML]{F9F9F9}Disadvantage and treat one group of users differently from another                                          \\
                                                   & Covert                                 & Hiding the influence mechanism from users                                                                   \\ \midrule
\multirow{2}{*}{Manipulation the information flow} & \cellcolor[HTML]{F9F9F9}Deceptive                              & \cellcolor[HTML]{F9F9F9}Induce false beliefs in users either through affirmative misstatements, misleading statements, or omissions \\
                                                   & Information Hiding                     & Obscure or delay the presentation of necessary information to users                                         \\ \bottomrule
\end{tabular}}
\end{table*}

\subsubsection{Disparate treatment}
The attributes from Mathur et al.~\cite{mathur2019cscw} describe most dark patterns that appear in most taxonomies, but do not cover a category of dark patterns that disadvantage and treat one group of users differently from another. These dark patterns are commonly found in games, as described by Zagal et al.~\cite{zagal2013dark}. For example, the Pay to Skip interface on gaming sites allows users with more resources to gain an edge over users who cannot afford to pay. Other interfaces that systematically disadvantage one group of users often operate covertly so that users may not know why they are shown a limited set of choices~\cite{hannak2014measuring}. To account for such interfaces we add disparate treatment as an attribute of dark patterns. Table~\ref{tab:dp} summarizes the complete set of attributes.

\subsection{Dark Patterns as a Thematic Research Area}
\label{sec:dark_patterns_themes}
The set of attributes described above exhibit commonalities. Dark patterns that are asymmetric, covert, or restrictive, or that involve disparate treatment, attempt to influence user decisions by \textbf{modifying the set of choices} available to users. Dark patterns that are deceptive or information hiding influence user decisions by \textbf{manipulating the information} that is available to to users. Ultimately, both of these themes reflect how dark patterns \textbf{modify the underlying choice architecture} for users.

As we show in Table~\ref{tab:dp-tax}, these themes---choice architecture, subdivided into user decision space and information flows---cohere the current dark patterns literature. We are able to provide a complete and cohesive descriptive taxonomy of dark patterns scholarship, from these themes, to attributes, to specific dark pattern user interface designs that are described in the prior work.

Importantly, \textit{themes} are what bind the current dark patterns scholarship together. We can only descriptively characterize the field at a high level of generality, and even then, it reflects a pair of related but distinct concepts. If we attempt to describe dark patterns with a greater degree of specificity---which would be essential for any actionable definition---the dark patterns literature immediately fragments further.

The structure of the dark patterns discourse has a parallel to Solove's seminal perspective on privacy discourse~\cite{solove2005taxonomy}. In the same way that Solove argued that privacy is a Wittgensteinian family of concepts that are distinct, but resemble each other, our descriptive deconstruction and reconstruction of the dark patterns literature reveals a set of interrelated thematic concerns. An important consequence of Solove's insight for privacy literature was that scholars could avoid being mired in distilling a unified definition of privacy, and instead focus on the normative considerations that motivate the different concepts at play.

We now turn to that task for dark patterns scholarship. Using our thematic observations about choice architecture as a starting point, we review related work in other disciplines. We then develop discrete normative perspectives that can help researchers more effectively articulate and study the problems that dark patterns pose for individuals and society.

\def\arraystretch{1.2}
\begin{footnotesize}
\begin{longtable}{@{}rlcccclcc@{}}
\caption{Types of dark pattern interfaces classified along how they modify the user's choice architecture.}
\label{tab:dp-tax}\\

\toprule
& & \multicolumn{7}{c}{\textbf{Choice Architecture}} \\ \cmidrule(l){3-9} 

\multicolumn{1}{c}{\textbf{Prior Work}} & \multicolumn{1}{c}{\textbf{Dark Pattern Type}} & \multicolumn{4}{c}{\textbf{Modify the Decision Space}} & & \multicolumn{2}{c}{\textbf{Manipulate the Info. Flow}} \\

\cmidrule(lr){3-6}
\cmidrule(l){8-9} 

& & \textbf{Asymme.} & \textbf{Restri.} & \textbf{Dis.\ Treat.} & \textbf{Covert} & & \textbf{Decept.} & \textbf{Info.\ Hiding} \\

\cmidrule(r){1-6}
\cmidrule(l){8-9} 
\endhead

\multirow{12}{*}{Brignull~\cite{brignull}} 

& \cellcolor[HTML]{F9F9F9}Bait and Switch & \cellcolor[HTML]{F9F9F9} & \cellcolor[HTML]{F9F9F9} & \cellcolor[HTML]{F9F9F9} & \cellcolor[HTML]{F9F9F9} & \cellcolor[HTML]{F9F9F9} & \cellcolor[HTML]{F9F9F9}\CIRCLE & \cellcolor[HTML]{F9F9F9} \\

& Confirmshaming & \CIRCLE &  &  &  &  &  &  \\

& \cellcolor[HTML]{F9F9F9}Disguised Ads & \cellcolor[HTML]{F9F9F9} & \cellcolor[HTML]{F9F9F9} & \cellcolor[HTML]{F9F9F9} & \cellcolor[HTML]{F9F9F9} & \cellcolor[HTML]{F9F9F9} & \cellcolor[HTML]{F9F9F9}\CIRCLE & \cellcolor[HTML]{F9F9F9}\\

& Forced Continuity &  &  &  &  &  & \RIGHTcircle & \CIRCLE \\

& \cellcolor[HTML]{F9F9F9}Friend Spam & \cellcolor[HTML]{F9F9F9} & \cellcolor[HTML]{F9F9F9} & \cellcolor[HTML]{F9F9F9} & \cellcolor[HTML]{F9F9F9} & \cellcolor[HTML]{F9F9F9} & \cellcolor[HTML]{F9F9F9}\CIRCLE & \cellcolor[HTML]{F9F9F9}\\

& Hidden Costs &  &  &  &  &  & \RIGHTcircle & \CIRCLE \\

& \cellcolor[HTML]{F9F9F9}Misdirection & \cellcolor[HTML]{F9F9F9} & \cellcolor[HTML]{F9F9F9} & \cellcolor[HTML]{F9F9F9} & \cellcolor[HTML]{F9F9F9}\CIRCLE & \cellcolor[HTML]{F9F9F9} & \cellcolor[HTML]{F9F9F9} & \cellcolor[HTML]{F9F9F9} \\

& Price Comp. Prevention &  & \CIRCLE &  &  &  &  & \CIRCLE \\

& \cellcolor[HTML]{F9F9F9}Privacy Zuckering & \cellcolor[HTML]{F9F9F9}\RIGHTcircle & \cellcolor[HTML]{F9F9F9}\RIGHTcircle & \cellcolor[HTML]{F9F9F9} & \cellcolor[HTML]{F9F9F9} & \cellcolor[HTML]{F9F9F9} & \cellcolor[HTML]{F9F9F9} & \cellcolor[HTML]{F9F9F9}\RIGHTcircle \\
 
& Roach Motel &  & \CIRCLE &  &  &  &  &  \\
                                           
& \cellcolor[HTML]{F9F9F9}Sneak into Basket & \cellcolor[HTML]{F9F9F9} & \cellcolor[HTML]{F9F9F9} & \cellcolor[HTML]{F9F9F9} & \cellcolor[HTML]{F9F9F9} & \cellcolor[HTML]{F9F9F9} & \cellcolor[HTML]{F9F9F9}\RIGHTcircle & \cellcolor[HTML]{F9F9F9}\CIRCLE \\

& Trick Questions & \CIRCLE &  &  & \CIRCLE &  &  &  \\

\cmidrule{2-9}

\multirow{9}{*}{Conti \& Sobiesk~\cite{conti2010malicious}}       & \cellcolor[HTML]{F9F9F9}Coercion & \cellcolor[HTML]{F9F9F9} & \cellcolor[HTML]{F9F9F9}\CIRCLE & \cellcolor[HTML]{F9F9F9} & \cellcolor[HTML]{F9F9F9} & \cellcolor[HTML]{F9F9F9} & \cellcolor[HTML]{F9F9F9} & \cellcolor[HTML]{F9F9F9} \\
 
 & Confusion & \CIRCLE &  &  & \CIRCLE &  &  &  \\
 
 & \cellcolor[HTML]{F9F9F9}Distraction & \cellcolor[HTML]{F9F9F9}\RIGHTcircle & \cellcolor[HTML]{F9F9F9} & \cellcolor[HTML]{F9F9F9} & \cellcolor[HTML]{F9F9F9}\CIRCLE & \cellcolor[HTML]{F9F9F9} & \cellcolor[HTML]{F9F9F9} & \cellcolor[HTML]{F9F9F9}\\

 & Forced Work & \RIGHTcircle & \RIGHTcircle & & & & & \\
 
 & \cellcolor[HTML]{F9F9F9}Interruption & \cellcolor[HTML]{F9F9F9}\RIGHTcircle & \cellcolor[HTML]{F9F9F9}\RIGHTcircle & \cellcolor[HTML]{F9F9F9} & \cellcolor[HTML]{F9F9F9} & \cellcolor[HTML]{F9F9F9} & \cellcolor[HTML]{F9F9F9} & \cellcolor[HTML]{F9F9F9}\\
 
 & Manipulating Navigation & \RIGHTcircle & &  & \RIGHTcircle & & & \\
 
 & \cellcolor[HTML]{F9F9F9}Obfuscation & \cellcolor[HTML]{F9F9F9}\RIGHTcircle & \cellcolor[HTML]{F9F9F9} & \cellcolor[HTML]{F9F9F9} & \cellcolor[HTML]{F9F9F9}\RIGHTcircle & \cellcolor[HTML]{F9F9F9} & \cellcolor[HTML]{F9F9F9} & \cellcolor[HTML]{F9F9F9} \\
 
 & Restricting Functionality & \RIGHTcircle & \RIGHTcircle & & & & & \\
 
 & \cellcolor[HTML]{F9F9F9}Trick & \cellcolor[HTML]{F9F9F9} & \cellcolor[HTML]{F9F9F9} & \cellcolor[HTML]{F9F9F9} & \cellcolor[HTML]{F9F9F9}\RIGHTcircle & \cellcolor[HTML]{F9F9F9} & \cellcolor[HTML]{F9F9F9}\RIGHTcircle & \cellcolor[HTML]{F9F9F9}\\
 
 \cmidrule{2-9}                                           
                                           
\multirow{7}{*}{Zagal et al.~\cite{zagal2013dark}} & Grinding & & \CIRCLE & & & & & \\
    & \cellcolor[HTML]{F9F9F9}Impersonation & \cellcolor[HTML]{F9F9F9} & \cellcolor[HTML]{F9F9F9} & \cellcolor[HTML]{F9F9F9} & \cellcolor[HTML]{F9F9F9} & \cellcolor[HTML]{F9F9F9} & \cellcolor[HTML]{F9F9F9}\CIRCLE & \cellcolor[HTML]{F9F9F9} \\
    
    & Monetized Rivalries & & & \CIRCLE & & & & \\
    
    & \cellcolor[HTML]{F9F9F9}Pay to Skip & \cellcolor[HTML]{F9F9F9} & \cellcolor[HTML]{F9F9F9} & \cellcolor[HTML]{F9F9F9}\CIRCLE & \cellcolor[HTML]{F9F9F9} & \cellcolor[HTML]{F9F9F9} & \cellcolor[HTML]{F9F9F9} & \cellcolor[HTML]{F9F9F9} \\
    
    & Playing by Appointment & & \CIRCLE & & & & & \\
    
    & \cellcolor[HTML]{F9F9F9}Pre-Delivered Content & \cellcolor[HTML]{F9F9F9} & \cellcolor[HTML]{F9F9F9} & \cellcolor[HTML]{F9F9F9}\CIRCLE & \cellcolor[HTML]{F9F9F9} & \cellcolor[HTML]{F9F9F9} & \cellcolor[HTML]{F9F9F9}\RIGHTcircle & \cellcolor[HTML]{F9F9F9}\CIRCLE \\
    
    & Social Pyramid Schemes & & \CIRCLE & & & & & \\ 
    
    \cmidrule{2-9}

\multirow{8}{*}{Greenberg et al.~\cite{greenberg-proxemics-2014}} & \cellcolor[HTML]{F9F9F9}Attention Grabber & \cellcolor[HTML]{F9F9F9}\RIGHTcircle & \cellcolor[HTML]{F9F9F9} & \cellcolor[HTML]{F9F9F9} & \cellcolor[HTML]{F9F9F9}\CIRCLE & \cellcolor[HTML]{F9F9F9} & \cellcolor[HTML]{F9F9F9} & \cellcolor[HTML]{F9F9F9} \\

& Bait and Switch & & &  & & & \CIRCLE & \\

& \cellcolor[HTML]{F9F9F9}Captive Audience & \cellcolor[HTML]{F9F9F9}\RIGHTcircle & \cellcolor[HTML]{F9F9F9}\CIRCLE & \cellcolor[HTML]{F9F9F9} & \cellcolor[HTML]{F9F9F9} & \cellcolor[HTML]{F9F9F9} & \cellcolor[HTML]{F9F9F9} & \cellcolor[HTML]{F9F9F9} \\

& Disguised Data Collection &  &  &  & \CIRCLE &  &  & \RIGHTcircle \\

& \cellcolor[HTML]{F9F9F9}Milk Factor & \cellcolor[HTML]{F9F9F9} & \cellcolor[HTML]{F9F9F9}\CIRCLE & \cellcolor[HTML]{F9F9F9} & \cellcolor[HTML]{F9F9F9} & \cellcolor[HTML]{F9F9F9} & \cellcolor[HTML]{F9F9F9} & \cellcolor[HTML]{F9F9F9}  \\

& Never Forget & & & &\CIRCLE & & & \\

& \cellcolor[HTML]{F9F9F9}Personal Info.\ Public & \cellcolor[HTML]{F9F9F9} & \cellcolor[HTML]{F9F9F9}\CIRCLE & \cellcolor[HTML]{F9F9F9} & \cellcolor[HTML]{F9F9F9} & \cellcolor[HTML]{F9F9F9} & \cellcolor[HTML]{F9F9F9} & \cellcolor[HTML]{F9F9F9} \\

& Unintended Relationships & & & & \CIRCLE & & & \\\cmidrule{2-9}
          
\multirow{7}{*}{B{\"o}sch et al.~\cite{bosch-tales-2016}} & \cellcolor[HTML]{F9F9F9}Address Book Leeching & \cellcolor[HTML]{F9F9F9} & \cellcolor[HTML]{F9F9F9} & \cellcolor[HTML]{F9F9F9} & \cellcolor[HTML]{F9F9F9} & \cellcolor[HTML]{F9F9F9} & \cellcolor[HTML]{F9F9F9}\CIRCLE & \cellcolor[HTML]{F9F9F9} \\

& Bad Defaults & \CIRCLE & & & & & & \\

& \cellcolor[HTML]{F9F9F9}Forced Registration & \cellcolor[HTML]{F9F9F9} & \cellcolor[HTML]{F9F9F9}\CIRCLE & \cellcolor[HTML]{F9F9F9} & \cellcolor[HTML]{F9F9F9} & \cellcolor[HTML]{F9F9F9} & \cellcolor[HTML]{F9F9F9} & \cellcolor[HTML]{F9F9F9} \\

& Hidden Legal.\ Stip.\ & & & & & & & \CIRCLE \\

& \cellcolor[HTML]{F9F9F9}Immortal Accounts & \cellcolor[HTML]{F9F9F9}\RIGHTcircle & \cellcolor[HTML]{F9F9F9}\RIGHTcircle & \cellcolor[HTML]{F9F9F9} & \cellcolor[HTML]{F9F9F9} & \cellcolor[HTML]{F9F9F9} & \cellcolor[HTML]{F9F9F9} & \cellcolor[HTML]{F9F9F9} \\

& Privacy Zuckering & \RIGHTcircle & \RIGHTcircle & & & & & \RIGHTcircle \\

& \cellcolor[HTML]{F9F9F9}Shadow User Profiles & \cellcolor[HTML]{F9F9F9} & \cellcolor[HTML]{F9F9F9}\CIRCLE & \cellcolor[HTML]{F9F9F9} & \cellcolor[HTML]{F9F9F9}\CIRCLE & \cellcolor[HTML]{F9F9F9} & \cellcolor[HTML]{F9F9F9} & \cellcolor[HTML]{F9F9F9} \\ \cmidrule{2-9}

\multirow{5}{*}{Gray et al.~\cite{gray-dark-patterns-2018}} & Forced Action & & \CIRCLE & & & & & \\

& \cellcolor[HTML]{F9F9F9}Interface Interference & \cellcolor[HTML]{F9F9F9}\RIGHTcircle & \cellcolor[HTML]{F9F9F9} & \cellcolor[HTML]{F9F9F9} & \cellcolor[HTML]{F9F9F9}\RIGHTcircle & \cellcolor[HTML]{F9F9F9} & \cellcolor[HTML]{F9F9F9} & \cellcolor[HTML]{F9F9F9}\RIGHTcircle \\

& Nagging & \CIRCLE & & & & & & \\

& \cellcolor[HTML]{F9F9F9}Obstruction & \cellcolor[HTML]{F9F9F9} & \cellcolor[HTML]{F9F9F9}\RIGHTcircle & \cellcolor[HTML]{F9F9F9} & \cellcolor[HTML]{F9F9F9}\RIGHTcircle & \cellcolor[HTML]{F9F9F9} & \cellcolor[HTML]{F9F9F9} & \cellcolor[HTML]{F9F9F9} \\

& Sneaking & & & & & & \RIGHTcircle & \CIRCLE \\ \cmidrule{2-9}

\multirow{5}{*}{NCC~\cite{frobrukerradet-deceived-2018}}
 & \cellcolor[HTML]{F9F9F9}Default Settings & \cellcolor[HTML]{F9F9F9}\CIRCLE & \cellcolor[HTML]{F9F9F9} & \cellcolor[HTML]{F9F9F9} & \cellcolor[HTML]{F9F9F9} & \cellcolor[HTML]{F9F9F9} & \cellcolor[HTML]{F9F9F9} & \cellcolor[HTML]{F9F9F9} \\

& Ease & \CIRCLE &  &  &  &  &  &  \\

& \cellcolor[HTML]{F9F9F9}Framing & \cellcolor[HTML]{F9F9F9}\CIRCLE & \cellcolor[HTML]{F9F9F9} & \cellcolor[HTML]{F9F9F9} & \cellcolor[HTML]{F9F9F9}\RIGHTcircle & \cellcolor[HTML]{F9F9F9} & \cellcolor[HTML]{F9F9F9} & \cellcolor[HTML]{F9F9F9}\\

& Rewards \& Punishment & \CIRCLE & \RIGHTcircle &  &  &  &  &  \\

& \cellcolor[HTML]{F9F9F9}Forced Action & \cellcolor[HTML]{F9F9F9} & \cellcolor[HTML]{F9F9F9}\CIRCLE & \cellcolor[HTML]{F9F9F9} & \cellcolor[HTML]{F9F9F9} & \cellcolor[HTML]{F9F9F9} & \cellcolor[HTML]{F9F9F9} & \cellcolor[HTML]{F9F9F9} \\\cmidrule{2-9}

\multirow{18}{*}{CNIL~\cite{cnil}}
 & Attention Diversion & \RIGHTcircle & & & \RIGHTcircle & & & \\
 & \cellcolor[HTML]{F9F9F9}Bait \& Change & \cellcolor[HTML]{F9F9F9} & \cellcolor[HTML]{F9F9F9} & \cellcolor[HTML]{F9F9F9} & \cellcolor[HTML]{F9F9F9} & \cellcolor[HTML]{F9F9F9} & \cellcolor[HTML]{F9F9F9}\CIRCLE & \cellcolor[HTML]{F9F9F9} \\
 & Blaming the Individual & \CIRCLE & & & & & & \\
 & \cellcolor[HTML]{F9F9F9}Camouflaged Advertising & \cellcolor[HTML]{F9F9F9} & \cellcolor[HTML]{F9F9F9} & \cellcolor[HTML]{F9F9F9} & \cellcolor[HTML]{F9F9F9} & \cellcolor[HTML]{F9F9F9} & \cellcolor[HTML]{F9F9F9}\CIRCLE & \cellcolor[HTML]{F9F9F9} \\
 & Chameleon Strategy & \RIGHTcircle &  & & & & & \CIRCLE \\
 & \cellcolor[HTML]{F9F9F9}Comparison Obfuscation & \cellcolor[HTML]{F9F9F9} & \cellcolor[HTML]{F9F9F9}\CIRCLE & \cellcolor[HTML]{F9F9F9} & \cellcolor[HTML]{F9F9F9} & \cellcolor[HTML]{F9F9F9} & \cellcolor[HTML]{F9F9F9} & \cellcolor[HTML]{F9F9F9}\CIRCLE \\
 & Default Sharing & \CIRCLE & & & & & & \\
 & \cellcolor[HTML]{F9F9F9}False Continuity & \cellcolor[HTML]{F9F9F9} & \cellcolor[HTML]{F9F9F9} & \cellcolor[HTML]{F9F9F9} & \cellcolor[HTML]{F9F9F9} & \cellcolor[HTML]{F9F9F9} & \cellcolor[HTML]{F9F9F9}\RIGHTcircle & \cellcolor[HTML]{F9F9F9}\CIRCLE \\
 & Hard to Adjust Privacy &\CIRCLE & & & & & & \\
 & \cellcolor[HTML]{F9F9F9}Impenetrable Wall & \cellcolor[HTML]{F9F9F9} & \cellcolor[HTML]{F9F9F9}\CIRCLE & \cellcolor[HTML]{F9F9F9} & \cellcolor[HTML]{F9F9F9} & \cellcolor[HTML]{F9F9F9} & \cellcolor[HTML]{F9F9F9} & \cellcolor[HTML]{F9F9F9} \\
 & Improving Experience & &  & &\RIGHTcircle & & & \RIGHTcircle \\
 & \cellcolor[HTML]{F9F9F9}Just You and Us & \cellcolor[HTML]{F9F9F9} & \cellcolor[HTML]{F9F9F9} & \cellcolor[HTML]{F9F9F9} & \cellcolor[HTML]{F9F9F9}\RIGHTcircle & \cellcolor[HTML]{F9F9F9} & \cellcolor[HTML]{F9F9F9} & \cellcolor[HTML]{F9F9F9}\RIGHTcircle \\
 & Last Minute Consent & & \RIGHTcircle & & \RIGHTcircle & & & \\
 & \cellcolor[HTML]{F9F9F9}Obfuscating Settings & \cellcolor[HTML]{F9F9F9}\CIRCLE & \cellcolor[HTML]{F9F9F9} & \cellcolor[HTML]{F9F9F9} & \cellcolor[HTML]{F9F9F9} & \cellcolor[HTML]{F9F9F9} & \cellcolor[HTML]{F9F9F9} & \cellcolor[HTML]{F9F9F9} \\
 & Repetitive Incentive & \RIGHTcircle & & & & & &\RIGHTcircle \\
 & \cellcolor[HTML]{F9F9F9}Safety Blackmail & \cellcolor[HTML]{F9F9F9} & \cellcolor[HTML]{F9F9F9} & \cellcolor[HTML]{F9F9F9} & \cellcolor[HTML]{F9F9F9} & \cellcolor[HTML]{F9F9F9} & \cellcolor[HTML]{F9F9F9}\CIRCLE & \cellcolor[HTML]{F9F9F9}\\
 & Trick Question & \CIRCLE & & & \CIRCLE & & & \\
 & \cellcolor[HTML]{F9F9F9}Wrong Signal & \cellcolor[HTML]{F9F9F9} & \cellcolor[HTML]{F9F9F9} & \cellcolor[HTML]{F9F9F9} & \cellcolor[HTML]{F9F9F9} & \cellcolor[HTML]{F9F9F9} & \cellcolor[HTML]{F9F9F9} & \cellcolor[HTML]{F9F9F9}\CIRCLE \\
 \cmidrule{2-9}

\multirow{6}{*}{Mathur et al.~\cite{mathur2019cscw}}
 & Forced Action &  & \CIRCLE &  &  &  &  &  \\

& \cellcolor[HTML]{F9F9F9}Misdirection & \cellcolor[HTML]{F9F9F9}\RIGHTcircle & \cellcolor[HTML]{F9F9F9} & \cellcolor[HTML]{F9F9F9} & \cellcolor[HTML]{F9F9F9}\RIGHTcircle & \cellcolor[HTML]{F9F9F9} & \cellcolor[HTML]{F9F9F9}\RIGHTcircle & \cellcolor[HTML]{F9F9F9} \\

& Obstruction &  & \CIRCLE &  &  &  &  & \RIGHTcircle \\

& \cellcolor[HTML]{F9F9F9}Scarcity & \cellcolor[HTML]{F9F9F9} & \cellcolor[HTML]{F9F9F9} & \cellcolor[HTML]{F9F9F9} & \cellcolor[HTML]{F9F9F9}\RIGHTcircle & \cellcolor[HTML]{F9F9F9} & \cellcolor[HTML]{F9F9F9}\RIGHTcircle & \cellcolor[HTML]{F9F9F9}\RIGHTcircle \\

& Sneaking &  &  &  &  &  & \RIGHTcircle & \CIRCLE \\

& \cellcolor[HTML]{F9F9F9}Social Proof & \cellcolor[HTML]{F9F9F9} & \cellcolor[HTML]{F9F9F9} & \cellcolor[HTML]{F9F9F9} & \cellcolor[HTML]{F9F9F9}\RIGHTcircle & \cellcolor[HTML]{F9F9F9} & \cellcolor[HTML]{F9F9F9}\RIGHTcircle & \cellcolor[HTML]{F9F9F9}\\\cmidrule{2-9}

Lacey \& Caudwell~\cite{cuteness} & Cuteness of Robots &  &  &  & \CIRCLE &  &  & \\\cmidrule{2-9}

\multirow{6}{*}{Gray et al.~\cite{gray-asshole}} & \cellcolor[HTML]{F9F9F9}Automating the User & \cellcolor[HTML]{F9F9F9} & \cellcolor[HTML]{F9F9F9}\CIRCLE & \cellcolor[HTML]{F9F9F9} & \cellcolor[HTML]{F9F9F9} & \cellcolor[HTML]{F9F9F9} & \cellcolor[HTML]{F9F9F9} & \cellcolor[HTML]{F9F9F9}\\

& Two-Faced & \CIRCLE &  &  & \CIRCLE &  & &  \\

& \cellcolor[HTML]{F9F9F9}Controlling & \cellcolor[HTML]{F9F9F9}\RIGHTcircle & \cellcolor[HTML]{F9F9F9}\RIGHTcircle & \cellcolor[HTML]{F9F9F9} & \cellcolor[HTML]{F9F9F9}\RIGHTcircle & \cellcolor[HTML]{F9F9F9} & \cellcolor[HTML]{F9F9F9} & \cellcolor[HTML]{F9F9F9}\\

& Entrapping &  & \CIRCLE &  & \CIRCLE &  &  &  \\

& \cellcolor[HTML]{F9F9F9}Nickle and Diming & \cellcolor[HTML]{F9F9F9}\CIRCLE & \cellcolor[HTML]{F9F9F9}\RIGHTcircle & \cellcolor[HTML]{F9F9F9} & \cellcolor[HTML]{F9F9F9} & \cellcolor[HTML]{F9F9F9} & \cellcolor[HTML]{F9F9F9}\RIGHTcircle & \cellcolor[HTML]{F9F9F9}\CIRCLE \\   

& Misrepresenting &  &  &  &  &  & \CIRCLE & \CIRCLE \\

\bottomrule

\end{longtable}

\mbox{}\hfill \CIRCLE \hspace{5pt}Required ``dark pattern'' attribute \hspace{10pt} \RIGHTcircle \hspace{5pt}Optional ``dark pattern'' attribute
\end{footnotesize}


%% file: sections/disciplines.tex
While the HCI community has begun to develop a dark patterns literature, other disciplines---including psychology, economics, philosophy, and law---have grappled with similar issues that result from modifications to users' choice architecture in contexts beyond user interfaces. Unlike HCI, however, these disciplines have deeply examined the normative implications that arise from modifying choice architecture. The HCI community should engage with the substantial scholarship in these disciplines to move beyond its current descriptive mode of thinking and to surface the normative implications of dark patterns. We synthesize representative scholarship from these disciplines below.

\subsection{Nudge and Sludge in Behavioral Economics} 
The field of behavioral economics is the study of psychology as it relates to the economic decision-making processes of individuals and organizations. Behavioral economists focus on the limits of human rationality, and how informational and cognitive capabilities influence these limits. In particular, they examine how human decision-making under uncertainty deviates from classical economic models because of how gains and losses are valued~\cite{tversky1979prospect}.

While the intellectual foundations for this approach were built in the late 1970s, the last 15 years has seen a surge in work that applies the lessons from behavioral economics to achieve various public policy goals. During this time, Thaler and Sunstein~\cite{thaler2009nudge} conceived of ``nudges,'' which they defined as ``private or public initiatives that steer people in particular directions but that also allow them to go their own way.'' Numerous types of nudges have found their way into public life. One common example is that setting the default choice to enroll users in a retirement savings plan boosts participation by overcoming inertia~\cite{johnson2003defaults}.

Initially, Thaler and Sunstein focused on the beneficial uses of nudges to further normative goals that increased individual and societal welfare~\cite{sunstein_2017}. More recently, they have ~\cite{Thaler431, sunstein2020sludge} denounced ``sludges''---nudges that induce excessive or unjustified friction and ``cost time or money; make life difficult to navigate; may be frustrating, stigmatizing or humiliating; and that might end up depriving people of access to important goods, opportunities and services.'' Ultimately, nudges or sludges are just tools to achieve (or detract from) particular policy goals and do not supply the normative rationale for what goals to adopt.

\subsection{Manipulation in Philosophy and Ethics} For decades, philosophy and ethics scholars have examined what it means to manipulate someone. Scholars agree that to manipulate an individual is to subvert their decision-making and to deny them authorship over their decisions~\cite{susser-online-manipulation, wilkinson2013nudging, Rudinow1978, wood2014coercion,frischmann_selinger_2018}. Manipulation is distinct from techniques like coercion and persuasion. Coercing someone is forcing or threatening them to take an action, while persuading someone involves a forthright appeal to their conscious decision-making capacities.

There is, however, a debate among these scholars on whether manipulation operates as a hidden force. Susser et al.~\cite{susser-online-manipulation}, for example, argue that manipulation is covert influence, meaning that the manner of the influence is not apparent to the individual being influenced. But others, like Wood~\cite{wood2014coercion} and Noggle~\cite{noggle1996manipulative}, argue that manipulation can be overt, citing examples where ``irresistible incentives'' lure individuals into selecting certain choices over others~\cite{Rudinow1978}.

From a normative standpoint, scholars have also raised numerous concerns about how manipulation can be exploitative, impoverish individuals, unfair, and---most important of all---undermine individual autonomy~\cite{susser-online-manipulation,frischmann_selinger_2018}. Manipulation can cause individuals to falsely believe they are making their own choices and are in control of their own destiny, and manipulation that targets individual vulnerabilities can amplify these effects.

\subsection{Market Manipulation in Law} Legal scholars Hanson and Kysar~\cite{hanson1999taking} were the first to describe a theory of market manipulation. They described market manipulation as the process of marketers and sellers exploiting individual non-rational behavior and cognitive biases for profit. For example, the framing effect bias explains variations in how users perceive a level of risk differently depending on the manner in which risk information is presented. Manufacturers exploit this cognitive bias in consumers to undermine the severity of product safety warnings to further their profits.

Calo~\cite{calo2013digital} argues that digital marketplaces and online stores make market manipulation even more effective because of the scale and sophistication with which these marketplaces can target users. For example, unlike brick-and-mortar stores, digital marketplaces can specifically learn about and optimize manipulative techniques to individual users, thereby increasing their average effect across the entire population. Calo and Rosenblat~\cite{calo2017taking} also argue that newer forms of marketplaces such as the sharing economy present at least as many---if not more---opportunities for digital market manipulation. Unlike traditional commerce marketplaces, the sharing economy marketplaces can manipulate both the sellers (e.g., Airbnb hosts) and the buyers (e.g., travelers).

In terms of the normative considerations of digital market manipulation, Calo~\cite{calo2013digital} argues that it can increase economic harm and reduce market efficiency. Market manipulation techniques that rely on personal data can lead to a loss of privacy for users and also undermine their autonomy.

%% file: sections/normative.tex
The previous section demonstrates how choice architecture problems in other disciplines have surfaced normative considerations that help analyze these problems. Unlike those disciplines, the dark patterns literature has largely focused on the descriptive aspects of dark patterns. Underlying these descriptive aspects, however, are a set of nascent normative concerns that attempt to explain \emph{why} dark patterns should concern us. We draw out these normative considerations from the literature and organize them into four normative lenses. We acknowledge that the considerations that these lenses offer will not appeal equally to everyone concerned about dark patterns. Rather, our goal here is to set the stage for researchers to develop a common language to discuss problematic practices.


\subsection{Individual Welfare}

The individual welfare normative lens views dark patterns on the basis of whether they diminish individual consumer welfare. Under this lens, a dark pattern is any interface that modifies the choice architecture to benefit the designer at the expense of the user's welfare. The dark patterns literature has highlighted the impact to individual welfare as the primary normative concern about dark patterns. When researchers argue that a dark pattern goes against the user's ``best interests,'' ``harms'' the user, or creates a ``negative experience'' for users~\cite{gray-asshole,gray-asshole,lewis-irresistible-apps-2014,frobrukerradet-deceived-2018, zagal2013dark,WALDMAN2020105}, they focus on some aspect of individual welfare. We discuss three kinds of individual welfare that can be diminished through dark patterns.

\subsubsection{Financial loss} The most straightforward welfare consequence to users is when a dark pattern causes them to suffer a financial loss. Shopping and travel websites have a wide variety of interfaces that push users into spending more than they originally anticipate. Examples of such interfaces include those that profit from users by adding products into shopping carts without users' consent (Sneak into Basket) and those that mislead users into believing they are signing up for a one-time offer or free trial when in reality they are signing up for recurring fees (Hidden Subscription). Across the web, advertisements that masquerade as non-advertising content (Disguised Ads) can also mislead users into purchasing products they may not have otherwise purchased~\cite{Mathuram}.

Concerns about financial loss echo throughout the dark patterns literature. Conti and Sobiesk~\cite{conti2010malicious} focus on ``malicious interfaces'' that ``are employed for a variety of reasons that are often linked to direct or indirect acquisition of revenue [such as] selling a product or service, increasing brand recognition...'' Zagal et al.~\cite{zagal2013dark} and Lewis~\cite{lewis-irresistible-apps-2014} argue that dark patterns cause ``negative experiences'' for users and tag those dark patterns that ``are a means for companies to extract extra money from users'' as monetary dark patterns. Brignull defined dark patterns as tricks that mislead users into ``buying or signing up for something [that they didn't mean to].''

\begin{figure}[t]
    \centering
    \setlength{\fboxsep}{1pt}%
\setlength{\fboxrule}{0.4pt}%
\fbox{
    \includegraphics[width=0.5\textwidth]{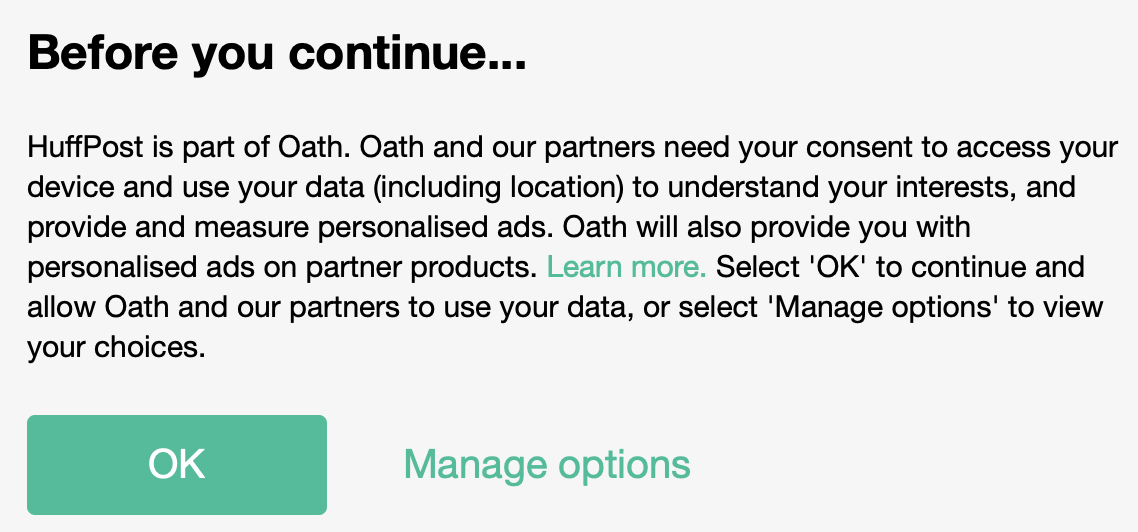}}
    \caption{A dark pattern in the cookie banner notice on HuffPost. The interface provides a positive option (``OK'') to consent to tracking and data collection but the negative option is hidden behind a sequence of seven screens starting from ``Manage Options.''}
    \label{fig:cookies}
        \Description{A dark pattern in the cookie banner notice on HuffPost. The interface provides a positive option (``OK'') to consent to tracking and data collection but the negative option is hidden behind a sequence of seven screens starting from ``Manage Options.''}
\end{figure}

\subsubsection{Invasion of privacy} Another welfare consequence to users is when a dark pattern undermines their privacy, for example, by undermining notice-and-choice privacy regimes.\footnote{Readers may also find Hartzog's \emph{Privacy’s Blueprint: The Battle to Control the Design of New Technologies}~\cite{hartzog2018privacy} as a valuable examination of how user interface design can impact privacy.} Examples of such interfaces include privacy-invasive defaults that expose user data (Bad Defaults), privacy-respecting choices that are hard to access (Privacy Zuckering, Interface Interference, and Obstruction), and the use of fear or other emotion-laden language to drive users away from making privacy-respecting choices (Blaming the Individual, Confirmshaming, and Framing).

The dark patterns literature is ripe with concerns about privacy as an aspect of individual welfare. Through their ``malicious interfaces'' framing, Conti and Sobiesk~\cite{conti2010malicious} argue that dark patterns extract revenue from the user by gathering their personal information. B{\"o}sch et al.~\cite{bosch-tales-2016} argue that ``[i]nstead of privacy-friendly solutions, [the proponents of dark patterns] aim for systems that purposefully and intentionally exploit their users’ privacy---for instance motivated by criminal reasons or financially exploitable business strategies.'' The NCC's report on dark patterns~\cite{frobrukerradet-deceived-2018} states that ``[t]he practice of misleading consumers into making certain choices, which may put their privacy at risk, is unethical and exploitative.'' Waldman~\cite{WALDMAN2020105} argues that ``[dark patterns] confuse users by asking questions in ways non experts cannot understand, they obfuscate by hiding interface elements that could help users protect their privacy, they require registration and associated disclosures in order to access functionality, and hide malicious behavior in the abyss of legalese privacy policies.'' A whole host of subsequent research considers how dark patterns diminish privacy~\cite{cnil, nouwens, circumvention, fomo, utz2019informed, machuletz2020multiple}. As an aside, it is important to note that while we take a welfarist view on privacy here, there are other views that describe privacy as a public good, a human right, or an aspect of individual autonomy.

\subsubsection{Cognitive burden} The last welfare consequence to users is when a dark pattern causes them to expend unnecessary time, energy, and attention. Imposing such a cognitive tax on users, especially when they have competing claims on their attention, can push users to select the easiest available choice, which suits the service at the expense of the user. Examples of such interfaces include those that obstruct users from canceling the online services they are subscribed to by requiring them to phone in only during certain hours (Hard to Cancel) and those that repeatedly prompt the user to accept certain choices (Nagging). In the case of Instacart---a popular grocery delivery application in the U.S.---the Nagging dark pattern continually beeped delivery workers' phones until they agreed to to accept lower-paying jobs~\cite{instacart}.

The concern about the cognitive burden of an interface has not been cited an explicit normative concern in the dark patterns literature. If the concern is raised, it is in the context of specific dark patterns such as Nagging, Hard to Cancel, and Hidden Legalese Stipulations (and other information overloading dark patterns). As one example, in the context of cookie consent dialogs, Soe et al.~\cite{circumvention} argue that cookie consent dialogs without a negative option to deny consent ``also introduce additional cognitive burden on the user.'' Figure~\ref{fig:cookies} shows once instance of a cookie consent dialog with a hard to exercise deny option.

\textbf{Evaluation}. The challenge for the individual consumer welfare perspective is that it is difficult to describe the baseline standard we should have in mind when examining potential dark patterns. Individual preferences and desires are difficult to measure for a variety of reasons, including because they are unstable, vary across individuals, and may require large-scale measurement. Critics also point out that all marketing involves shaping user preferences to varying degrees and so this lens says little about how we can distinguish benign or tolerable practices from ones that deserve sanction. Another challenge is that this perspective does not account for the dynamics of competitive markets where users can punish bad actors for mistreating them by abandoning them or imposing some reputational cost. Finally, this perspective does not account for potential welfare benefits that may flow from allowing companies to shape the decision space or manipulate information flow, or the costs of imposing liability for all manners of marketing practices.

\begin{figure}[t]
    \centering
    \setlength{\fboxsep}{1pt}%
\setlength{\fboxrule}{0.4pt}%
\fbox{
    \includegraphics[width=0.5\textwidth]{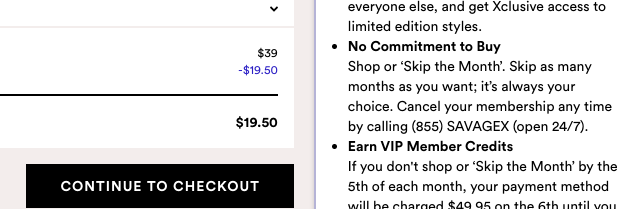}}
    \caption{The Obstruction (Hard to Cancel) dark pattern on Savage X Fenty. While users can sign up for a membership online, they can only cancel their membership by placing a phone call to customer care.}
    \label{fig:obstruction}
            \Description{The Obstruction (Hard to Cancel) dark pattern on Savage X Fenty. While users can sign up for a membership online, they can only cancel their membership by placing a phone call to customer care.}
\end{figure}

\subsection{Collective Welfare}

The collective welfare lens takes a broader normative perspective on consumer welfare compared to the individual welfare lens, and views dark patterns on the basis of whether they diminish collective welfare. Under this lens, a dark pattern is any interface that modifies the choice architecture to benefit the designer at the expense of collective welfare. Unlike individual welfare, which has been the primary normative focus of the dark patterns literature, collective welfare has received limited attention. We discuss four kinds of collective welfare associated with society and markets that can be diminished through dark patterns:

\subsubsection{Competition} An efficient and competitive marketplace enables innovation, results in lower prices for products, and helps consumers align with products that match their preferences. However, dominant providers in particular can abuse their position of monopoly power and diminish competition by making it look like consumers independently selected their product rather than from a series of dark patterns that disfavor offers from competitors. Dark patterns such as a pre-selected checkboxes (Preselection) and use of emotions and fear (Framing) can allow providers to tie-in unrelated products to unfairly gain market share despite having an inferior tied product. Dark Patterns from the Obstruction category (Figure~\ref{fig:obstruction}) can assist providers in creating high switching costs for users to potentially superior competitors while simultaneously creating a high barrier of entry for others. A prominent example is the Microsoft antitrust case,\footnote{United States v.Microsoft Corp., 253 F.3d 34, 65 (D.C. Cir. 2001).} where Microsoft used the Obstruction dark pattern to prevent users from uninstalling Internet Explorer like other programs in Windows 98.

The dark patterns literature has only provided limited commentary about competition concerns emerging from dark patterns. An exception is the work of Day and Stemler~\cite{darkanticompetitive}, which raises the primary concern that dark patterns ``erode users' ability to act rationally, which empowers platforms to extract wealth and build market power without doing so on the merits.'' Day and Stemler argue that dark patterns harm competition because ``when the design is meant to generate switching costs without an offsetting benefit to users,'' and that the key in making this judgement should be ``whether the design was meant to enhance addiction and manipulate usage while providing consumers with a qualitatively worse product.''

\subsubsection{Price transparency} Price transparency enables consumers to make informed decisions and thus creates an efficient marketplace. Providers do not, however, always have the incentive to compete on prices. For example, oil companies were using a variety of names and descriptions for the same product until the FTC stepped in in 1971 to issue rules requiring standardized octane ratings for fuel to prevent consumer confusion.\footnote{Statement of Basis and Purpose, Posting of Minimum Octane Numbers on Gasoline Dispensing Pumps, 36 Fed. Reg. 23871, 23882 (1971).} A more contemporary instance is the use of different product names and attributes to sell the same product, as exemplified by the flat screen TV market.

Dark patterns hide the true costs of products from consumers and prevent them from comparison shopping. Dark patterns like Hidden Costs and Price Comparison Prevention impede consumers from making informed decisions. As a result, not only does opaque pricing cause potential financial loses for consumers---they might have selected a cheaper alternative if they were able to compare and contrast competing prices---but it also reduces competition in the marketplace. Economists argue that under certain conditions providers gain no benefit from exposing the opaque prices of their competitors~\cite{gabaix2006shrouded}, and instead, they find it more profitable to employ and copy the same practices.

\subsubsection{Trust in the market} Dark patterns can undermine consumer trust in markets and hurt companies who engage in legitimate and honest practices. As these patterns proliferate, users who become aware of them may become skeptical of and resistant to interface elements that \textit{look} like dark patterns. For example, studies have shown that users have become resistant and wise to Scarcity dark patterns~\cite{wise-nudge}. But this increased skepticism in users may lead them to miss out on genuine deals from honest retailers and hurt the business of those companies. Through their study, Luguri and Strahilevitz~\cite{luguri2019shining} demonstrate that web services making use of egregious dark patterns can suffer repercussions, as users are more likely to report negative emotions after interacting with them. While this might suggest dark patterns are ultimately self-defeating, the problem is that the bad actor may not be the only one paying the price.

\subsubsection{Unanticipated societal consequences} More generally, design choices can also lead to consequences that are not originally anticipated by the designer. For example, interfaces employed by companies like Facebook make users turn over private data to allow the company to amass large databases with detailed profiles about individuals that it can use for advertising. But that information in turn can also be repurposed to create products that can undermine societal values. A prominent example here is how Cambridge Analytica harvested Facebook user data and initiated a political disinformation campaign to impact the U.S.\ presidential election in 2016. As a result, a collective consumer welfare perspective requires evaluating potential unintended or side effects of design choices. 

\textbf{Evaluation}. This attention to collective welfare consequences also explains how while certain dark patterns may not present a problem to a particular user's  welfare, they can impose a cumulative cost on other members of society that is intolerable. An example is the Social Pyramid Schemes in gaming apps that encourage users to allow the service to spam their friends and acquaintances or nudge them to sign up. The users may understand and accept the trade-off, but there is a significant cost to society as a whole from tolerating such practices.  The collective welfare approach depends on having a mechanism to balance the different costs and benefits that result from design choices. Should designs deliver the great good for the greatest number of people? Or should design protect the interests of more vulnerable communities of users? 

\subsection{Regulatory Objectives}

The regulatory objectives lens uses democratically created rules and standards to view when dark patterns cause individual and collective harms such as diminishing the individual's financial welfare and undermining fair market competition respectively. Under this lens, a dark pattern is any interface that modifies the choice architecture to interfere with or undermine specific regulatory objectives. The lens takes the operating legal regime as a given in making this judgement about an interface.

Generally speaking, the underlying regulations that define the standards that any interface needs to meet do not have to be described with a great deal of specificity. For example, interfaces that are deceptive because they omit material information or rely on false information can run afoul of well established consumer protection laws prohibiting such practices. As a result, regulators do not have a single standard to decide what constitutes a problematic practice---they tend to take a case-by-case approach to address distinct problems. Sometimes authorities also pass specific laws to regulate certain practices like the EU did with the General Data Protection Regulation (GDPR)~\cite{gdpr} to handle privacy.

In the U.S., the Federal Trade Commission (FTC) prominently uses it authority under Section 5(a) of the Federal Trade Commission Act to determine whether a company has acted unfairly or deceptively.\footnote{15 U.S.C. \textsection{} 45(a)(1).} The FTC evaluates whether a practice is deceptive from the perspective of a ``reasonable consumer''---meaning, does the practice leave a false or misleading impression on a typical consumer? The Commission also has the authority to prohibit non-deceptive practices that have the effect of treating consumers unfairly. The FTC has a long history of taking action to ``halt some form of seller behavior that unreasonably creates or takes advantage of an obstacle to the free exercise of consumer decision-making,'' even without proof of financial loss~\cite{FTChalt}. In such cases, the FTC determines whether the act or practice ``causes or is likely to cause substantial injury to consumers which is not reasonably avoidable by consumers themselves and not outweighed by countervailing benefits to consumers or to competition.''\footnote{15 U.S.C. \textsection{} 45(n).} State attorneys general have analogous powers under state law to prohibit unfair or deceptive practices.

In recent years, however, the FTC has focused on only providing remedies for ``significant injuries'' like financial losses. The FTC's reasoning is based on the prevailing ``efficient market hypothesis'' among U.S. regulators, which posits that because users value choice and independent decision-making, companies will compete to provide products and services that do not take advantage of consumers~\cite{pitofsky}. Consequently, companies that fail to match that demand will not succeed in this competitive marketplace. Industry self-regulatory standards to police harmful conduct can also supplement this reliance on the markets. But if there are instances of market failures, the authorities turn to public and private enforcement actions to reign in the companies' conduct.

Authorities can also pass regulations that protect users by including mandatory disclosures about material facts or a cooling off period to allow users the time to evaluate information in a less pressured setting. There is considerable variation in how different legal regimes approach the problem. In the EU, for example, policymakers are predisposed to pass detailed laws and regulations defining permissible and impermissible conduct, whereas U.S.\ regulators tend to take a principles-based approach.

Aside from unfair and deceptive standards, legislators in the U.S. have proposed the DETOUR Act~\cite{detour-act} to regulate broad swaths of dark patterns. The DETOUR Act proposes a standard for dark patterns and makes it unlawful for a large online operator to ``to design, modify, or manipulate a user interface with the purpose or substantial effect of obscuring, subverting, or impairing user autonomy, decision-making, or choice to obtain consent or user data.'' Among other things, the bill requires online sites to be more transparent about how they use design features to influence user behavior. Relatedly, the recently passed CPRA~\cite{caballot} makes it illegal for businesses for obtain certain privacy consent through the use of dark patterns, which the law defines as ``user interfaces designed or manipulated with the substantial effect of subverting or impairing user autonomy, decision-making, or choice, as further defined by regulation.'' 

The dark patterns literature has offered some initial commentary on whether dark patterns are lawful. Conti and Sobiesk~\cite{conti2010malicious} note that ``malicious interfaces'' can ``obfuscat[e] legally mandated but undesirable information from the user.'' Mathur et al.~\cite{mathur2019cscw} argue that the deceptive dark patterns they discovered are unambiguously unlawful under Section 5 of the FTC Act and various state laws in the U.S. Luguri and Strahilevitz~\cite{luguri2019shining} offer a detailed commentary about the lawfulness of dark patterns under Section 5 along with other federal frameworks. However, only recently has the literature begun evaluating dark patterns against specific regulatory objectives. The majority of this discussion has resulted from studies~\cite{circumvention, nouwens, utz2019informed, machuletz2020multiple} examining whether the consent management platforms that have emerged on websites as a result of the GDPR are compliant with the law. Nouwens et al.~\cite{nouwens} consider consent management platforms compliant with the GDPR if the consent they aim to collect through their interface is explicit, is easy to accept or deny, and contains no preselected checkboxes. Failing to meet these criteria makes the consent interface a potentially unlawful dark pattern. Soe et al.~\cite{circumvention} similarly provide a list of eight criteria that would make a consent management platform interface a dark pattern under the GDPR.

\textbf{Evaluation}.  The regulatory objectives perspective on dark patterns is more instrumental then normative---the most forceful normative arguments for implementing regulatory objectives are typically the normative arguments for those objectives, rather than compliance with regulation in the abstract. This perspective does not inherently advance a normative argument about why we should care financial losses, privacy harms, or cognitive burdens, beyond noting whether the law directs us to care about those values. This perspective does come with a significant advantage, though: fashioning regulation into measurable metrics for empirical research is usually much easier than adapting normative principles to research.

\subsection{Individual Autonomy}

In contrast to the welfare and regulatory perspectives we have discussed so far, the individual autonomy lens is a rights-based lens. Autonomy is the normative value that users have the right to act on their own reasons when making decisions~\cite{frischmann_selinger_2018, susser-online-manipulation}. The autonomy perspective shares some common ground with the individual welfare perspective, but it does not make a cost-benefit decision to determine if an interface is problematic. Under this lens, a dark pattern is a user interface that undermines individual decision-making. A dark pattern that infringes on individual autonomy might modify choice architecture in a way that causes users to make choices that they would not have otherwise selected absent the modified choice architecture.\footnote{This standard for a dark pattern that infringes on autonomy interests is strongly linked to the welfare perspective too, since the standard prioritizes outcomes consistent with consumer preferences. From an individual welfare perspective, consistency with individual preferences is desirable because preferences are a possible measure of welfare. From an individual autonomy perspective, consistency with individual preferences is desirable because preferences are a gauge of autonomous decision-making.} Alternatively, a dark pattern might deny a user choice, obscure available choices, or burden the exercise of choice.

The vast majority of dark patterns attempt to undermine individual autonomy. One set of such dark patterns include those that enable addiction. Online services have incentives to maximize user engagement with their platform that is at odds with user autonomy. Dark patterns that cause digital addiction can lead to a variety of negative consequences including psychological (e.g., poor concentration), and physical (e.g., sleep disturbance) harms~\cite{addiction}. In fact, Sch{\"u}ll observed that online services have created techniques such as the infinite scroll user interface, or the YouTube autoplay feature in line with those developed by the gambling industry---many of which employ a variable-rewards schedule approach to keep users engaged~\cite{schull2014addiction,lewis-irresistible-apps-2014}. Lootboxes---variable reward containers that give users advantages over others in video games---have also come under scrutiny because of their ability to continually engage kids and teen gamers~\cite{ftc-lootbox}.

Concerns about autonomy echo through nearly the entire dark patterns literature. Brignull~\cite{brignull} defined dark patterns as ``tricks that make you do things that you didn't mean to...'' The NCC's report~\cite{frobrukerradet-deceived-2018} argues that dark patterns ``deprive [users] of their agency.'' The DETOUR Act~\cite{detour-act} and the CPRA~\cite{caballot} both describe dark patterns as ``subverting or impairing user autonomy.'' Westin and Chiasson~\cite{fomo} argue that dark patterns ``trick and manipulate users into taking an action they would've other-wise been unlikely to take.'' Further, with regards to determining whether dark patterns are ethical or not, Westin and Chiasson argue that ``it is the loss of individual autonomy promoted by dark patterns that we should carefully consider.''

\textbf{Evaluation}. The autonomy perspective is appealing, because it captures many of the concerns articulated in prior dark pattern definitions and taxonomies. One drawback of the autonomy perspective is that it sweeps broadly, dubbing all interfaces that interfere with decision making as dark patterns regardless of whether the outcome of that decision benefits the individual or society. Another challenge is that the autonomy perspective depends on a highly idealized version of human action that does not account for lived realities, where decisions are made with limited information and scarce resources for deliberation. Yet another challenge is line drawing: how do we distinguish between permissible burdens on autonomy (e.g., persuasive advertising that changes behavior) from dark patterns that violate autonomy? And, relatedly, how can we measure whether and the extent to which a user interface violates individual autonomy?

%% file: sections/measure.tex
The four normative lenses---individual welfare, collective welfare, regulatory objectives, and individual autonomy---offer us different perspectives with which we can analyze dark patterns. The perspectives do not, however, tell us how to evaluate user interfaces in practice. In this section, we describe how HCI researchers can use established empirical methods to analyze dark patterns through these four normative lenses. Grounding empirical analysis in normative considerations enables researchers to surface the problematic aspects of dark patterns in a principled and defensible manner. Our goal is to encourage the HCI community to move beyond ad hoc and descriptive labels for dark patterns and focus future research on assessing how, exactly, dark patterns interact with specific normative concerns.

To set the stage, we first provide a brief description of various kinds of HCI research methods. We describe the breadth of information each method is capable of collecting and how that information can be subsequently analyzed. Next, we illustrate how the methods can be applied to study the normative lenses we previously described. We close the section by applying the methods in a case study to analyze the Trick Questions dark pattern.

\subsection{Empirical Research Methods in HCI}

The HCI community has championed the use of methods that elicit users' attitudes and behaviors to help design and evaluate user interfaces. Each one of these methods has its own merits and drawbacks, and---depending on how it is ultimately applied---provides researchers with a different set of insights about user interfaces.  Note that this description of research methods is not meant to be comprehensive. Instead, we refer interested readers to standard HCI textbooks~\cite{olson2014ways,lazar2017research}.

One set of methods allows researchers to study user experience with a particular design by directly interacting with or observing users. This set of methods includes surveys, interviews, focus groups, and ethnography. Surveys are a structured set of questions that can be deployed to gather responses from a large sample of users. Interviews on the other hand are typically less structured and allow researchers to conduct an in-depth and back-and-forth discussion with users. Interviews enable a deeper inquiry because of their flexibility, but unlike surveys they are harder to scale. In focus groups, a researcher directs a discussion with a group of users to elicit individual and collective feedback about a design. Ethnography involves passively observe the behavior of individuals and groups in their own social contexts.

Lab and field studies are settings that allow researchers to examine actual user interaction with a user interface. In a lab study, researchers typically invite users into the lab and observe them interacting with a particular design. Researchers then ask users about their experience on interaction cues of interest, or after the study is complete in the form of a cognitive walkthrough. Lab measurement methods can be particularly rigorous, such as gaze tracking systems that enable researchers to understand user attention to specific components of user interfaces. An important drawback of the lab setting is that it often has limited external validity, because of sampling, response, and other biases. Lab insights may not generalize to how users interact with a design in the real world.

Instead, researchers may use field deployments to study user interaction \emph{in situ}. Field deployments can include diary studies and experience sampling. Both these methods require users to complete a prompt of questions immediately after an event of interest when using a particular design, outside of their interaction with researchers. These deployments can be particularly useful in studying the experience of sub-populations that may be otherwise hard to reach. Field deployments with measurement panels can provide valuable insights into how users behave in the presence of a particular user interface.

Experiments---the bedrock of the scientific method---enable generating cause-and-effect relationships. Typically in an experiment, researchers randomly assign users to the control or treatment conditions, and then measure some quantity of interest across those conditions. Statistical significance in the measure between the conditions reveals the effect of the treatment. As with the previous set of methods, experiments may be conducted in the lab or in the real world. Lab experiments provide a controlled setting, and they minimize the influence of external factors on the experiment. Field experiments on the other hand enable inquiry in a more realistic and everyday setting, and also help in recruiting a relatively more diverse sample of participants. Online marketplaces like Amazon Mechanical Turk and Prolific are popular venues to conduct such large-scale field experiments.

\subsection{Applying HCI Methods to Analyze Dark Patterns}

Each of the lenses we described in Section~\ref{sec:norm} lends itself to a different set of measurement methods, metrics, and goals. In the following sections, we describe how the measurement methods above can be used to analyze dark patterns. As the dark patterns literature has only recently begun employing user studies~\cite{conti2010malicious, nouwens,maier2020dark,utz2019informed,machuletz2020multiple,luguri2019shining,dig} to analyze user interfaces, we discuss these studies in the context of how they align with normative lenses.

\subsubsection{What should we measure under each normative lens?\\}

\textbf{Individual welfare:} 
The first normative lens describes how dark patterns diminish individual welfare. From this perspective, the goal of any measurement should be to measure the change in individual welfare that results from a dark pattern. If the individual welfare of concern is financial value, then a study should measure the financial loss suffered by users. An experiment comparing the dark pattern interface to a neutral design could measure the average monetary amount lost by users because of the dark pattern. The experiment conducted by Luguri and Strahilevitz~\cite{luguri2019shining} is one relevant example. The study measured the rate at which participants accepted a hypothetical insurance offering under the absence and presence of dark patterns, then calculated the amount of money lost by those who were subject to the dark pattern. On the other hand, if the individual welfare of concern is privacy, then a study should measure a welfarist conception of privacy. It could measure the financial value of the data that was disclosed due to the dark pattern, or it could measure the fraction of users whose preferences were not satisfied because of the dark pattern. These could be measured using a survey, field study, or experiment comparing the dark pattern interface to a neutral design. Finally, if the individual welfare of concern is cognitive burden, then a study could measure the cognitive burden imposed on users. A survey or experiment could employ various neurophysiological~\cite{fredericks2005investigation} or stated measures of cognitive load to compare dark patterns and neutral designs~\cite{sweller2018measuring}. The survey conducted by Conti and Sobiesk~\cite{conti2010malicious} is an example of a user study that measured the level of frustration users felt with various kinds of dark patterns.

\textbf{Collective welfare:} The second normative lens describes how dark patterns diminish collective welfare. From this perspective, the goal of any measurement should be to measure the change in collective welfare that results from a dark pattern. If the collective welfare of concern is competition, then a study should measure the reduction in the competitiveness of the marketplace. Competition could be measured in several different ways. As one example, a study could first develop an understanding of individual behavior under a dark pattern when compared to a neutral design. It could then induce the broader effect of this behavior on competition in the marketplace, such as by using simulations and modeling. On the other hand, if the collective welfare of concern is marketplace trust, then a study should measure the loss of trust among the various market participants. A study could use a survey to measure consumers' perceptions of the trustworthiness of the marketplace in the absence and presence of a dark pattern.

\textbf{Regulatory objectives:} The third normative lens describes how dark patterns interfere with regulations. Under this lens, the goal of any measurement should be to assess whether a dark pattern complies with relevant regulation. For example, the GDPR requires that consent must be ``freely given, specific, informed and unambiguous indication of the data subject's wishes by which he or she, by a statement or by a clear affirmative action, signifies agreement to the processing of personal data relating to him or her.'' A study could measure compliance with the GDPR by examining whether users are able to express their privacy preferences and provide consent free of coercion in the presence of a dark pattern. The online experiments conducted by Nouwens et al.~\cite{nouwens}, Utz et al.~\cite{utz2019informed}, and Machuletz et al.~\cite{machuletz2020multiple} are relevant examples. All three studies analyzed whether designs commonly used by consent management platforms are compliant with the law. Among other metrics, each study measured the fraction of users who consented to have their data processed by a website under the presence of a dark pattern. The experiment conducted by Nouwens et al.~\cite{nouwens} additionally measured the fraction of users whose preferences were satisfied by the consent interface. This additional dimension measures privacy from an individual welfare perspective.

\textbf{Individual autonomy:} The fourth normative lens describes how dark patterns undermine individual autonomy. From this perspective, the goal of any measurement should be to measure the extent to which a dark pattern interferes with users' ability to make independent decisions. Comparing a dark pattern to a neutral design, a study could use a survey or experiment to measure whether users perceive a loss of autonomy, whether users are aware of alternative choices, or whether users are unable to exercise their preferred choices. Again, the online experiments conducted by Utz et al.~\cite{utz2019informed} and Machuletz et al.~\cite{machuletz2020multiple} are relevant examples. In measuring compliance with the GDPR, both experiments additionally measured autonomy. Machuletz et al.~\cite{machuletz2020multiple} measured whether users found it difficult to express certain choices through the consent interface. Utz et al.~\cite{utz2019informed} measured the mental models that users had about the consent interface, including awareness of the set of choices offered by the interface.

\subsubsection{What is an appropriate baseline for measurement?\\}
Some of the measurements described in the previous section can be established on their own---without needing a baseline design to compare against. From the the regulatory objectives perspective, for example, measuring whether a dark pattern complies with a regulation might only require analyzing whether it complies with a clearly established rule or standard. Consider the FTC's endorsement guidelines, which regulate the disclosure of social media ads (the Disguised Ad dark pattern). The guidelines state that an undisclosed ad is problematic ``if it misleads a \emph{significant minority} of consumers.'' If a measurement reveals that a significant minority---for some threshold---has been misled, then the undisclosed ad does not comply with the guidelines. There is no need to evaluate the ad's design against a hypothetical alternative ad. As another example, consider the individual welfare and autonomy perspectives. A study might deem a design a dark pattern if a sufficient proportion of users, above some threshold, do not have their preferences or expectations satisfied by the user interface. The threshold could be a minority of users (as in the FTC's endorsement guidelines), or it could be a simple majority, or it could be a supermajority. Setting the threshold would require normative judgment, but again, would not require a comparison to alternative designs.

Another way to study a user interface without selecting a baseline is to compare the design to the space of nearly identical designs. Consider, for example, a consent form that involves difficult to understand prompts (the Trick Questions dark pattern). A study could compare the form to alternative versions of the form with slight differences in phrasing and presentation. If the initial form is an outlier in comparison to the space of similar forms (e.g., when comparing the proportion of users who agree or the proportion of users who have their preferences satisfied), that could be interpreted as evidence that the form is a dark pattern.

Many of the measurements described above, however, require a  baseline alternative user interface for comparison.\footnote{User interface designs that include multiple possible dark patterns pose an additional challenge. One approach would be to implement multiple baseline designs for comparison, examining each of the possible dark patterns separately. Another approach would be to combine the possible dark patterns for treatment and control purposes.} In comparative research designs like experiments, envisioning this neutral control user interface---a design not containing the dark pattern---is an important consideration. There are important parallels here with how regulators have approached deceptive advertising. Some past enforcement actions have involved comparing a concerning ad to a ``tombstone'' control ad~\cite{craswell1996compared}, which omits the potentially deceptive content.

So, how should HCI researchers pick an appropriate control design for dark pattern studies? We foresee at least three different options. First, the appropriate baseline may be specified by regulation itself, by what users reasonably prefer, or by what users might generally expect of the interface. Second, the appropriate baseline may be derived from the space of designs that are similar to the concerning user interface but that do not contain the possible dark pattern. In many instances, it will be possible to produce a slight variant of a user interface that simply omits the possible dark pattern. Third, the appropriate baseline may be selected from the space of similar designs used by other online services for similar choices. For example, if one online service's sign-up flow includes a possible dark pattern component, but a competing service's sign-up flow does not include that component, then the competitor's user interface may be an appropriate baseline for comparison.

\subsubsection{What is an appropriate threshold for identifying a dark pattern?\\}
Regardless of whether an empirical study involves a baseline user interface, if a study aims to characterize designs as dark patterns (or not), it requires setting a threshold for what constitutes a dark pattern. Luguri and Strahilevitz~\cite{luguri2019shining} propose one possible threshold: if the uptake rate associated with a user interface is over doubled in comparison to an alternative user interface, then the user interface is a dark pattern. The rationale, they argue, is that this test means a user's uptake is more likely than not attributable to a dark pattern. While this is one possible approach, as noted in the prior section, there are many others. The threshold could be absolute, or relative via subtraction, or relative as a ratio, among other options. The threshold could be modest, or it could be a high bar for labeling a user interface a dark pattern. Setting the threshold requires careful normative justification.

\subsubsection{Who is affected by the dark pattern?\\}
Any user study ultimately needs to be representative of the users that are targeted by the dark pattern. This could be the general public or special sub-populations like children, older adults, or low-income individuals. A more difficult challenge concerns evaluating the effect of a dark pattern on third parties---users who do not interact with the interface directly. For example, addictive online services like games can have an detrimental effect on not just on the user, but also on their family and friends. Similarly, a number of growth hacking dark patterns rely on users giving easy access to their social network that might seem relatively trivial to them, but has a cumulative negative impact of dealing with unwanted communications that can be quite expensive for their friends and family. However, even if a dark pattern is targeted at the general public, it can have disproportionate impacts on certain sub-populations. Study designs should account for conducting such analyses.

\subsubsection{What role can non-quantitative methods play?\\}
While we have so far discussed quantitative methods, other HCI methods have an important role to play in analyzing dark patterns through the lenses we describe. Interviews and focus groups, like the ones conducted by Maier and Harr~\cite{maier2020dark}, could be used to analyze dark patterns that affect sub-populations that might be hard to reach otherwise. Future work could, for example, use ethnographic methods and observe designers to understand how user interfaces become dark patterns during development. Results from this work could shed light on multiple normative considerations.

\subsubsection{How are dark patterns different from marketing?\\}

Some commentators have questioned whether dark patterns are merely aggressive marketing practices, the sort of sharp dealing that has long existed (and been tolerated) for offline businesses~\cite{cric1,cric2}. Researchers have responded by noting possible key differences between dark patterns and traditional marketing, including low cost, large scale, and unprecedented sophistication~\cite{mathur2019cscw}. Researchers have especially highlighted the role of A/B testing and machine learning, which can easily generate and evaluate user interfaces in ways that may be problematic for users~\cite{mathur2019cscw, narayanan2020dark}. Future work could involve measurements that are grounded in a normative perspective and specifically address how these factors compare to traditional marketing. For example, studies could compare online dark patterns to analogous offline marketing designs, or could simulate how machine learning might naturally converge on user interface designs that undermine autonomy or harm users.

\begin{figure}[t]
    \centering\setlength{\fboxsep}{1pt}%
\setlength{\fboxrule}{0.4pt}%
\fbox{
    \includegraphics[width=0.5\textwidth]{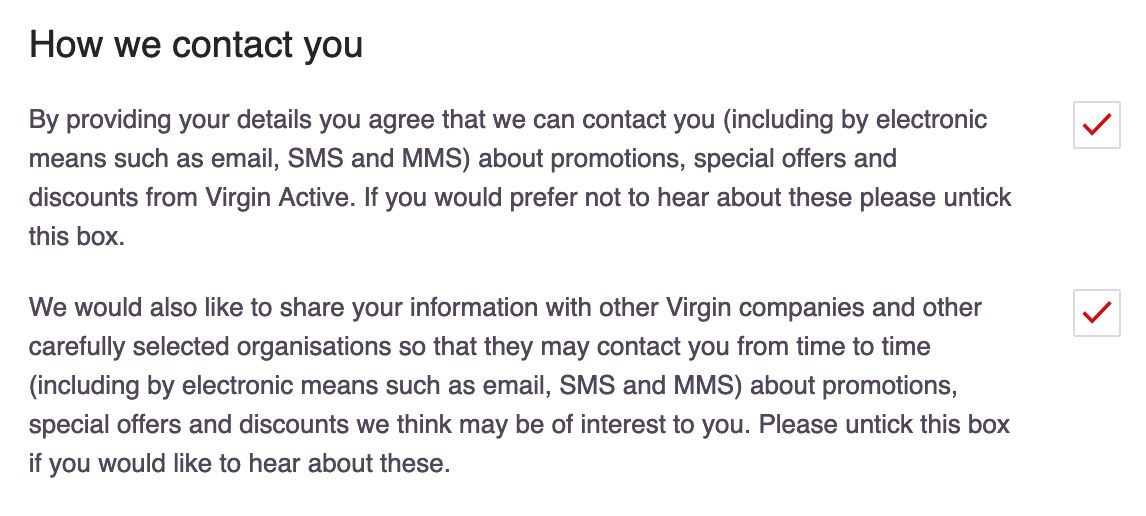}}
    \caption{The Trick Questions dark pattern on Virgin Active. Opting out of marketing and promotions requires unticking the first checkbox and leaving the second checkbox ticked.}
    \label{fig:tq}
                \Description{The Trick Questions dark pattern on Virgin Active. Opting out of marketing and promotions requires unticking the first checkbox and leaving the second checkbox ticked.}
\end{figure}

\subsection{Case Study: Trick Questions Dark Pattern}

In the last section, we discussed how a wide variety of HCI methods could be used to analyze dark patterns. We now present a case study to concretely demonstrate how researchers might apply those methods to evaluate a user interface in a manner that is grounded in a normative perspective. The case study is based on the Trick Questions dark pattern, in which confusing language steers users into making certain choices. Figure~\ref{fig:tq} illustrates one instance of a Trick Questions user interface.

Each of the four normative lenses provides a different way of approaching a Trick Questions user interface. Starting with the individual welfare lens, a researcher might ask: does this user interface undermine individual welfare? The researcher might answer that question by conducting a randomized controlled lab experiment, examining the extent to which the interface respects user preferences. The researcher could generate an alternative baseline design---say, a user interface with unambiguous explanation of the user's choices---then compare the choices users make with the original design and the alternative design. If the choices with the alternative design are significantly closer to user preferences than the choices with the original design, the researcher could interpret that as evidence that the original design undermined individual welfare as a Trick Questions dark pattern. Alternatively, where user choices have a monetary component (e.g., making a payment), the researcher could compare the costs imposed on users between the original and alternative designs. Again, if there is a significant difference, the researcher could interpret that as evidence of a dark pattern from the individual welfare perspective.

Now consider the collective welfare lens, and assume the researcher's primary collective welfare concern is market trust. The researcher might again use a randomized controlled laboratory experiment, but measuring how users self-report trust in the relevant online services market before and after completing the experiment task. If there is a significant drop in trust for the initial design but not the alternative design, that could be interpreted as evidence of a Trick Questions dark pattern that undermines collective welfare.


Turning to the regulatory objective lens, a researcher might seek to measure whether the user interface complies with a specific regulatory obligation, such as the GDPR's requirement that consent be ``freely given, specific, informed and unambiguous.'' The researcher could conduct a survey, in which participants receive a screenshot of the user interface and are asked questions about what each option presented in the interface does. If only a small proportion of users correctly understand what the privacy options in the user interface do, that could be interpreted as evidence that the user interface is a Trick Questions dark pattern from the regulatory perspective.

Finally, consider the individual autonomy lens. The researcher might seek to understand whether users are aware that the user interface provides them with options, such that they can exercise independent decision-making. The researcher could again use a survey method, but focusing on whether participants even notice the available options in the user interface. If only a small proportion of users recognize that there are options available, that could be interpreted as evidence that the user interface involves Trick Questions from the individual autonomy perspective.

The example research approaches we have just sketched are far from exhaustive, and in many respects the approaches are relatively unsophisticated and naive. Our goal with this case study is to show that when approaching dark patterns measurement research, scholars can and should articulate an informed decision about the normative perspectives they are applying, then articulate a further informed decision about the research methods they will use to shed light on those normative considerations. 


%% file: sections/conclusion.tex
Our primary goal in this paper is to advance the dark patterns literature in the HCI research community. Our community can benefit from scholarship in other disciplines (Section \ref{sec:disciplines}), which we develop into a set of normative perspectives for assessing dark patterns (Section \ref{sec:norm}). We then show how empirical dark patterns research methods can be grounded in these specific normative perspectives (Section \ref{sec:hmeasure}).

We encourage our colleagues to pursue this path in future work. Focusing on the normative foundations of dark patterns scholarship will motivate research problems and methods, and it will lend analytical rigor to understanding how dark patterns affect individuals and society. This approach will also benefit HCI practitioners by furthering our community's quest to develop ethical standards for user interface design. And by converging on principled methods for evaluating dark patterns, we can provide practitioners with a new toolkit for critically examining their own practices. This effort will also aid the HCI community in coming to terms with its responsibility for the societal consequences of its work. In particular, we have seen how insights about human interactions can used to exploit users. Our paper is not immune from that problem, but we trust that empowering researchers and regulators to examine the risks of dark patterns far outweighs the risk of describing mechanisms that a bad actor might seek to exploit.

Finally, grounding the measurement of dark patterns in normative perspectives will be invaluable for policymakers. Dark patterns have attracted serious legislative and regulatory attention in the U.S.\ and the EU. But policymakers have struggled with how to respond, since the issue is so amorphous and under-conceptualized. As a result, government actions on dark patterns have thus far remained limited to traditional consumer protection and privacy law. The HCI community can break through this logjam by developing metrics and methods for regulatory agencies to deploy.

%% file: main.bbl

\begin{thebibliography}{62}


\ifx \showCODEN    \undefined \def \showCODEN     #1{\unskip}     \fi
\ifx \showDOI      \undefined \def \showDOI       #1{#1}\fi
\ifx \showISBNx    \undefined \def \showISBNx     #1{\unskip}     \fi
\ifx \showISBNxiii \undefined \def \showISBNxiii  #1{\unskip}     \fi
\ifx \showISSN     \undefined \def \showISSN      #1{\unskip}     \fi
\ifx \showLCCN     \undefined \def \showLCCN      #1{\unskip}     \fi
\ifx \shownote     \undefined \def \shownote      #1{#1}          \fi
\ifx \showarticletitle \undefined \def \showarticletitle #1{#1}   \fi
\ifx \showURL      \undefined \def \showURL       {\relax}        \fi
\providecommand\bibfield[2]{#2}
\providecommand\bibinfo[2]{#2}
\providecommand\natexlab[1]{#1}
\providecommand\showeprint[2][]{arXiv:#2}

\bibitem[\protect\citeauthoryear{Beales}{Beales}{2003}]%
        {FTChalt}
\bibfield{author}{\bibinfo{person}{J.~Howard Beales}.}
  \bibinfo{year}{2003}\natexlab{}.
\newblock \bibinfo{title}{The FTC's Use of Unfairness Authority: Its Rise,
  Fall, and Resurrection}.
\newblock
  \bibinfo{howpublished}{\url{https://www.ftc.gov/public-statements/2003/05/ftcs-use-unfairness-authority-its-rise-fall-and-resurrection}}.
\newblock


\bibitem[\protect\citeauthoryear{Berthon, Pitt, and Campbell}{Berthon
  et~al\mbox{.}}{2019}]%
        {addiction}
\bibfield{author}{\bibinfo{person}{Pierre Berthon}, \bibinfo{person}{Leyland
  Pitt}, {and} \bibinfo{person}{Colin Campbell}.}
  \bibinfo{year}{2019}\natexlab{}.
\newblock \showarticletitle{Addictive De-Vices: A Public Policy Analysis of
  Sources and Solutions to Digital Addiction}.
\newblock \bibinfo{journal}{\emph{Journal of Public Policy \& Marketing}}
  \bibinfo{volume}{38}, \bibinfo{number}{4} (\bibinfo{year}{2019}),
  \bibinfo{pages}{451--468}.
\newblock
\urldef\tempurl%
\url{https://doi.org/10.1177/0743915619859852}
\showDOI{\tempurl}
\showeprint{https://doi.org/10.1177/0743915619859852}


\bibitem[\protect\citeauthoryear{Brignull}{Brignull}{2018}]%
        {brignull}
\bibfield{author}{\bibinfo{person}{Harry Brignull}.}
  \bibinfo{year}{2018}\natexlab{}.
\newblock \bibinfo{title}{Dark Patterns}.
\newblock \bibinfo{howpublished}{\url{https://darkpatterns.org/}}.
\newblock
\newblock
\shownote{Accessed September 9, 2020.}


\bibitem[\protect\citeauthoryear{Bösch, Erb, Kargl, Kopp, and
  Pfattheicher}{Bösch et~al\mbox{.}}{2016}]%
        {bosch-tales-2016}
\bibfield{author}{\bibinfo{person}{Christoph Bösch}, \bibinfo{person}{Benjamin
  Erb}, \bibinfo{person}{Frank Kargl}, \bibinfo{person}{Henning Kopp}, {and}
  \bibinfo{person}{Stefan Pfattheicher}.} \bibinfo{year}{01 Oct.
  2016}\natexlab{}.
\newblock \showarticletitle{Tales from the Dark Side: Privacy Dark Strategies
  and Privacy Dark Patterns}.
\newblock \bibinfo{journal}{\emph{Proceedings on Privacy Enhancing
  Technologies}} \bibinfo{volume}{2016}, \bibinfo{number}{4} (\bibinfo{year}{01
  Oct. 2016}), \bibinfo{pages}{237--254}.
\newblock
\urldef\tempurl%
\url{https://doi.org/10.1515/popets-2016-0038}
\showDOI{\tempurl}


\bibitem[\protect\citeauthoryear{Calo}{Calo}{2013}]%
        {calo2013digital}
\bibfield{author}{\bibinfo{person}{Ryan Calo}.}
  \bibinfo{year}{2013}\natexlab{}.
\newblock \showarticletitle{Digital market manipulation}.
\newblock \bibinfo{journal}{\emph{Geo. Wash. L. Rev.}}  \bibinfo{volume}{82}
  (\bibinfo{year}{2013}), \bibinfo{pages}{995}.
\newblock


\bibitem[\protect\citeauthoryear{Calo and Rosenblat}{Calo and
  Rosenblat}{2017}]%
        {calo2017taking}
\bibfield{author}{\bibinfo{person}{Ryan Calo} {and} \bibinfo{person}{Alex
  Rosenblat}.} \bibinfo{year}{2017}\natexlab{}.
\newblock \showarticletitle{The taking economy: Uber, information, and power}.
\newblock \bibinfo{journal}{\emph{Colum. L. Rev.}}  \bibinfo{volume}{117}
  (\bibinfo{year}{2017}), \bibinfo{pages}{1623}.
\newblock


\bibitem[\protect\citeauthoryear{Commission}{Commission}{2019}]%
        {cric2}
\bibfield{author}{\bibinfo{person}{Federal~Trade Commission}.}
  \bibinfo{year}{2019}\natexlab{}.
\newblock \bibinfo{title}{Competition and Consumer Protection in the 21st
  Century}.
\newblock
  \bibinfo{howpublished}{\url{https://www.ftc.gov/system/files/documents/public_events/1418273/ftc_hearings_session_12_transcript_day_1_4-9-19.pdf}}.
\newblock


\bibitem[\protect\citeauthoryear{Competition and Authority}{Competition and
  Authority}{2020}]%
        {ukreport}
\bibfield{author}{\bibinfo{person}{Competition} {and} \bibinfo{person}{Markets
  Authority}.} \bibinfo{year}{2020}\natexlab{}.
\newblock \bibinfo{title}{Online platforms and digital advertising market
  study}.
\newblock
  \bibinfo{howpublished}{\url{https://assets.publishing.service.gov.uk/media/5efc3faae90e075c4e144c69/Appendix_Y_-_Fairness_by_Design_Final_Version_v.8.pdf}}.
\newblock
\newblock
\shownote{Accessed September 9, 2020.}


\bibitem[\protect\citeauthoryear{Conti and Sobiesk}{Conti and Sobiesk}{2010}]%
        {conti2010malicious}
\bibfield{author}{\bibinfo{person}{Gregory Conti} {and} \bibinfo{person}{Edward
  Sobiesk}.} \bibinfo{year}{2010}\natexlab{}.
\newblock \showarticletitle{Malicious Interface Design: Exploiting the User}.
  In \bibinfo{booktitle}{\emph{Proceedings of the 19th International Conference
  on World Wide Web}} (Raleigh, North Carolina, USA)
  \emph{(\bibinfo{series}{WWW '10})}. \bibinfo{publisher}{Association for
  Computing Machinery}, \bibinfo{address}{New York, NY, USA},
  \bibinfo{pages}{271–280}.
\newblock
\showISBNx{9781605587998}
\urldef\tempurl%
\url{https://doi.org/10.1145/1772690.1772719}
\showDOI{\tempurl}


\bibitem[\protect\citeauthoryear{Craswell}{Craswell}{1996}]%
        {craswell1996compared}
\bibfield{author}{\bibinfo{person}{Richard Craswell}.}
  \bibinfo{year}{1996}\natexlab{}.
\newblock \showarticletitle{Compared to What-The Use of Control Ads in
  Deceptive Advertising Litigation}.
\newblock \bibinfo{journal}{\emph{Antitrust LJ}}  \bibinfo{volume}{65}
  (\bibinfo{year}{1996}), \bibinfo{pages}{757}.
\newblock


\bibitem[\protect\citeauthoryear{Day and Stemler}{Day and Stemler}{2020}]%
        {darkanticompetitive}
\bibfield{author}{\bibinfo{person}{Gregory Day} {and} \bibinfo{person}{Abbey
  Stemler}.} \bibinfo{year}{2020}\natexlab{}.
\newblock \showarticletitle{Are Dark Patterns Anticompetitive?}
\newblock \bibinfo{journal}{\emph{Alabama Law Review}}  \bibinfo{volume}{72}
  (\bibinfo{year}{2020}), \bibinfo{pages}{1--45}.
\newblock


\bibitem[\protect\citeauthoryear{Di~Geronimo, Braz, Fregnan, Palomba, and
  Bacchelli}{Di~Geronimo et~al\mbox{.}}{2020}]%
        {dig}
\bibfield{author}{\bibinfo{person}{Linda Di~Geronimo}, \bibinfo{person}{Larissa
  Braz}, \bibinfo{person}{Enrico Fregnan}, \bibinfo{person}{Fabio Palomba},
  {and} \bibinfo{person}{Alberto Bacchelli}.} \bibinfo{year}{2020}\natexlab{}.
\newblock \showarticletitle{UI Dark Patterns and Where to Find Them: A Study on
  Mobile Applications and User Perception}. In
  \bibinfo{booktitle}{\emph{Proceedings of the 2020 CHI Conference on Human
  Factors in Computing Systems}} (Honolulu, HI, USA)
  \emph{(\bibinfo{series}{CHI '20})}. \bibinfo{publisher}{Association for
  Computing Machinery}, \bibinfo{address}{New York, NY, USA},
  \bibinfo{pages}{1--14}.
\newblock
\showISBNx{9781450367080}
\urldef\tempurl%
\url{https://doi.org/10.1145/3313831.3376600}
\showDOI{\tempurl}


\bibitem[\protect\citeauthoryear{Eidelson}{Eidelson}{2019}]%
        {instacart}
\bibfield{author}{\bibinfo{person}{Josh Eidelson}.}
  \bibinfo{year}{2019}\natexlab{}.
\newblock \bibinfo{title}{Instacart Hounds Workers to Take Jobs That Aren’t
  Worth It}.
\newblock
  \bibinfo{howpublished}{\url{https://www.bloomberg.com/news/articles/2019-07-15/instacart-hounds-workers-to-take-jobs-that-aren-t-worth-it}}.
\newblock
\newblock
\shownote{Accessed September 9, 2020.}


\bibitem[\protect\citeauthoryear{{European Parliament and Council of European
  Union}}{{European Parliament and Council of European Union}}{2018}]%
        {gdpr}
\bibfield{author}{\bibinfo{person}{{European Parliament and Council of European
  Union}}.} \bibinfo{year}{2018}\natexlab{}.
\newblock \bibinfo{title}{Consent under the GDPR: valid, freely given,
  specific, informed and active consent}.
\newblock
  \bibinfo{howpublished}{\url{https://www.i-scoop.eu/gdpr/consent-gdpr/}}.
\newblock
\newblock
\shownote{Accessed September 9, 2020.}


\bibitem[\protect\citeauthoryear{Fogg}{Fogg}{1999}]%
        {fogg1999persuasive}
\bibfield{author}{\bibinfo{person}{Brian~J Fogg}.}
  \bibinfo{year}{1999}\natexlab{}.
\newblock \showarticletitle{Persuasive technologie301398}.
\newblock \bibinfo{journal}{\emph{Commun. ACM}} \bibinfo{volume}{42},
  \bibinfo{number}{5} (\bibinfo{year}{1999}), \bibinfo{pages}{26--29}.
\newblock


\bibitem[\protect\citeauthoryear{Fredericks, Choi, Hart, Butt, and
  Mital}{Fredericks et~al\mbox{.}}{2005}]%
        {fredericks2005investigation}
\bibfield{author}{\bibinfo{person}{Tycho~K Fredericks}, \bibinfo{person}{Sang~D
  Choi}, \bibinfo{person}{Jason Hart}, \bibinfo{person}{Steven~E Butt}, {and}
  \bibinfo{person}{Anil Mital}.} \bibinfo{year}{2005}\natexlab{}.
\newblock \showarticletitle{An investigation of myocardial aerobic capacity as
  a measure of both physical and cognitive workloads}.
\newblock \bibinfo{journal}{\emph{International Journal of Industrial
  Ergonomics}} \bibinfo{volume}{35}, \bibinfo{number}{12}
  (\bibinfo{year}{2005}), \bibinfo{pages}{1097--1107}.
\newblock


\bibitem[\protect\citeauthoryear{Frischmann and Selinger}{Frischmann and
  Selinger}{2018}]%
        {frischmann_selinger_2018}
\bibfield{author}{\bibinfo{person}{Brett Frischmann} {and}
  \bibinfo{person}{Evan Selinger}.} \bibinfo{year}{2018}\natexlab{}.
\newblock \bibinfo{booktitle}{\emph{Re-Engineering Humanity}}.
\newblock \bibinfo{publisher}{Cambridge University Press}.
\newblock
\urldef\tempurl%
\url{https://doi.org/10.1017/9781316544846}
\showDOI{\tempurl}


\bibitem[\protect\citeauthoryear{Frobrukerr\r{a}det}{Frobrukerr\r{a}det}{2018}]%
        {frobrukerradet-deceived-2018}
\bibfield{author}{\bibinfo{person}{Frobrukerr\r{a}det}.}
  \bibinfo{year}{2018}\natexlab{}.
\newblock \bibinfo{title}{Deceived by design: How tech companies use dark
  patterns to discourage us from exercising our rights to privacy}.
\newblock
\newblock


\bibitem[\protect\citeauthoryear{Gabaix and Laibson}{Gabaix and
  Laibson}{2006}]%
        {gabaix2006shrouded}
\bibfield{author}{\bibinfo{person}{Xavier Gabaix} {and} \bibinfo{person}{David
  Laibson}.} \bibinfo{year}{2006}\natexlab{}.
\newblock \showarticletitle{Shrouded attributes, consumer myopia, and
  information suppression in competitive markets}.
\newblock \bibinfo{journal}{\emph{The Quarterly Journal of Economics}}
  \bibinfo{volume}{121}, \bibinfo{number}{2} (\bibinfo{year}{2006}),
  \bibinfo{pages}{505--540}.
\newblock


\bibitem[\protect\citeauthoryear{Gray, Chivukula, and Lee}{Gray
  et~al\mbox{.}}{2020}]%
        {gray-asshole}
\bibfield{author}{\bibinfo{person}{Colin~M. Gray}, \bibinfo{person}{Shruthi~Sai
  Chivukula}, {and} \bibinfo{person}{Ahreum Lee}.}
  \bibinfo{year}{2020}\natexlab{}.
\newblock \showarticletitle{What Kind of Work Do ``Asshole Designers'' Create?
  Describing Properties of Ethical Concern on Reddit}. In
  \bibinfo{booktitle}{\emph{Proceedings of the 2020 ACM Designing Interactive
  Systems Conference}} (Eindhoven, Netherlands) \emph{(\bibinfo{series}{DIS
  '20})}. \bibinfo{publisher}{Association for Computing Machinery},
  \bibinfo{address}{New York, NY, USA}, \bibinfo{pages}{61–73}.
\newblock
\showISBNx{9781450369749}
\urldef\tempurl%
\url{https://doi.org/10.1145/3357236.3395486}
\showDOI{\tempurl}


\bibitem[\protect\citeauthoryear{Gray, Kou, Battles, Hoggatt, and Toombs}{Gray
  et~al\mbox{.}}{2018}]%
        {gray-dark-patterns-2018}
\bibfield{author}{\bibinfo{person}{Colin~M. Gray}, \bibinfo{person}{Yubo Kou},
  \bibinfo{person}{Bryan Battles}, \bibinfo{person}{Joseph Hoggatt}, {and}
  \bibinfo{person}{Austin~L. Toombs}.} \bibinfo{year}{2018}\natexlab{}.
\newblock \showarticletitle{The Dark (Patterns) Side of UX Design}. In
  \bibinfo{booktitle}{\emph{Proceedings of the 2018 CHI Conference on Human
  Factors in Computing Systems}} (Montreal QC, Canada)
  \emph{(\bibinfo{series}{CHI '18})}. \bibinfo{publisher}{ACM},
  \bibinfo{address}{New York, NY, USA}, Article \bibinfo{articleno}{534},
  \bibinfo{numpages}{14}~pages.
\newblock
\showISBNx{978-1-4503-5620-6}
\urldef\tempurl%
\url{https://doi.org/10.1145/3173574.3174108}
\showDOI{\tempurl}


\bibitem[\protect\citeauthoryear{Greenberg, Boring, Vermeulen, and
  Dostal}{Greenberg et~al\mbox{.}}{2014}]%
        {greenberg-proxemics-2014}
\bibfield{author}{\bibinfo{person}{Saul Greenberg}, \bibinfo{person}{Sebastian
  Boring}, \bibinfo{person}{Jo Vermeulen}, {and} \bibinfo{person}{Jakub
  Dostal}.} \bibinfo{year}{2014}\natexlab{}.
\newblock \showarticletitle{Dark Patterns in Proxemic Interactions: A Critical
  Perspective}. In \bibinfo{booktitle}{\emph{Proceedings of the 2014 Conference
  on Designing Interactive Systems}} (Vancouver, BC, Canada)
  \emph{(\bibinfo{series}{DIS '14})}. \bibinfo{publisher}{ACM},
  \bibinfo{address}{New York, NY, USA}, \bibinfo{pages}{523--532}.
\newblock
\showISBNx{978-1-4503-2902-6}
\urldef\tempurl%
\url{https://doi.org/10.1145/2598510.2598541}
\showDOI{\tempurl}


\bibitem[\protect\citeauthoryear{Hannak, Soeller, Lazer, Mislove, and
  Wilson}{Hannak et~al\mbox{.}}{2014}]%
        {hannak2014measuring}
\bibfield{author}{\bibinfo{person}{Aniko Hannak}, \bibinfo{person}{Gary
  Soeller}, \bibinfo{person}{David Lazer}, \bibinfo{person}{Alan Mislove},
  {and} \bibinfo{person}{Christo Wilson}.} \bibinfo{year}{2014}\natexlab{}.
\newblock \showarticletitle{Measuring Price Discrimination and Steering on
  E-Commerce Web Sites}. In \bibinfo{booktitle}{\emph{Proceedings of the 2014
  Conference on Internet Measurement Conference}} (Vancouver, BC, Canada)
  \emph{(\bibinfo{series}{IMC '14})}. \bibinfo{publisher}{Association for
  Computing Machinery}, \bibinfo{address}{New York, NY, USA},
  \bibinfo{pages}{305–318}.
\newblock
\showISBNx{9781450332132}
\urldef\tempurl%
\url{https://doi.org/10.1145/2663716.2663744}
\showDOI{\tempurl}


\bibitem[\protect\citeauthoryear{Hanson and Kysar}{Hanson and Kysar}{1999}]%
        {hanson1999taking}
\bibfield{author}{\bibinfo{person}{Jon~D Hanson} {and}
  \bibinfo{person}{Douglas~A Kysar}.} \bibinfo{year}{1999}\natexlab{}.
\newblock \showarticletitle{Taking behavioralism seriously: The problem of
  market manipulation}.
\newblock \bibinfo{journal}{\emph{NYUL Rev.}}  \bibinfo{volume}{74}
  (\bibinfo{year}{1999}), \bibinfo{pages}{630}.
\newblock


\bibitem[\protect\citeauthoryear{Hartzog}{Hartzog}{2018}]%
        {hartzog2018privacy}
\bibfield{author}{\bibinfo{person}{Woodrow Hartzog}.}
  \bibinfo{year}{2018}\natexlab{}.
\newblock \bibinfo{booktitle}{\emph{Privacy’s blueprint: the battle to
  control the design of new technologies}}.
\newblock \bibinfo{publisher}{Harvard University Press}.
\newblock


\bibitem[\protect\citeauthoryear{Hebert}{Hebert}{2019}]%
        {ftc-lootbox}
\bibfield{author}{\bibinfo{person}{Amy Hebert}.}
  \bibinfo{year}{2019}\natexlab{}.
\newblock \bibinfo{title}{Video games, loot boxes, and your money}.
\newblock
  \bibinfo{howpublished}{\url{https://www.consumer.ftc.gov/blog/2019/09/video-games-loot-boxes-and-your-money}}.
\newblock
\newblock
\shownote{Accessed September 9, 2020.}


\bibitem[\protect\citeauthoryear{Hurwitz}{Hurwitz}{2020}]%
        {cric1}
\bibfield{author}{\bibinfo{person}{Justin Hurwitz}.}
  \bibinfo{year}{2020}\natexlab{}.
\newblock \showarticletitle{Designing a Pattern, Darkly}.
\newblock \bibinfo{journal}{\emph{NCJL \& Tech.}}  \bibinfo{volume}{22}
  (\bibinfo{year}{2020}), \bibinfo{pages}{57}.
\newblock


\bibitem[\protect\citeauthoryear{Johnson and Goldstein}{Johnson and
  Goldstein}{2003}]%
        {johnson2003defaults}
\bibfield{author}{\bibinfo{person}{Eric~J Johnson} {and}
  \bibinfo{person}{Daniel Goldstein}.} \bibinfo{year}{2003}\natexlab{}.
\newblock \bibinfo{title}{Do defaults save lives?}
\newblock
\newblock


\bibitem[\protect\citeauthoryear{Lacey and Caudwell}{Lacey and
  Caudwell}{2019}]%
        {cuteness}
\bibfield{author}{\bibinfo{person}{Cherie Lacey} {and}
  \bibinfo{person}{Catherine Caudwell}.} \bibinfo{year}{2019}\natexlab{}.
\newblock \showarticletitle{Cuteness as a 'Dark Pattern' in Home Robots}. In
  \bibinfo{booktitle}{\emph{Proceedings of the 14th ACM/IEEE International
  Conference on Human-Robot Interaction}} \emph{(\bibinfo{series}{HRI '19})}.
  \bibinfo{publisher}{IEEE Press}, \bibinfo{address}{Daegu, Republic of Korea},
  \bibinfo{pages}{374–381}.
\newblock
\showISBNx{9781538685556}


\bibitem[\protect\citeauthoryear{Lazar, Feng, and Hochheiser}{Lazar
  et~al\mbox{.}}{2017}]%
        {lazar2017research}
\bibfield{author}{\bibinfo{person}{Jonathan Lazar},
  \bibinfo{person}{Jinjuan~Heidi Feng}, {and} \bibinfo{person}{Harry
  Hochheiser}.} \bibinfo{year}{2017}\natexlab{}.
\newblock \bibinfo{booktitle}{\emph{Research methods in human-computer
  interaction}}.
\newblock \bibinfo{publisher}{Morgan Kaufmann}, \bibinfo{address}{Boston}.
\newblock


\bibitem[\protect\citeauthoryear{Lewis}{Lewis}{2014}]%
        {lewis-irresistible-apps-2014}
\bibfield{author}{\bibinfo{person}{Chris Lewis}.}
  \bibinfo{year}{2014}\natexlab{}.
\newblock \bibinfo{booktitle}{\emph{Irresistible Apps: Motivational Design
  Patterns for Apps, Games, and Web-based Communities} (\bibinfo{edition}{1st}
  ed.)}.
\newblock \bibinfo{publisher}{Apress}, \bibinfo{address}{Berkely, CA, USA}.
\newblock
\showISBNx{1430264217, 9781430264217}


\bibitem[\protect\citeauthoryear{Luguri and Strahilevitz}{Luguri and
  Strahilevitz}{2020}]%
        {luguri2019shining}
\bibfield{author}{\bibinfo{person}{Jamie Luguri} {and} \bibinfo{person}{Lior
  Strahilevitz}.} \bibinfo{year}{2020}\natexlab{}.
\newblock \showarticletitle{Shining a Light on Dark Patterns}.
\newblock \bibinfo{journal}{\emph{Journal of Legal Analysis}}
  \bibinfo{volume}{12} (\bibinfo{year}{2020}), \bibinfo{pages}{Forthcoming}.
\newblock


\bibitem[\protect\citeauthoryear{Machuletz and B{\"o}hme}{Machuletz and
  B{\"o}hme}{2020}]%
        {machuletz2020multiple}
\bibfield{author}{\bibinfo{person}{Dominique Machuletz} {and}
  \bibinfo{person}{Rainer B{\"o}hme}.} \bibinfo{year}{2020}\natexlab{}.
\newblock \showarticletitle{Multiple purposes, multiple problems: A user study
  of consent dialogs after GDPR}.
\newblock \bibinfo{journal}{\emph{Proceedings on Privacy Enhancing
  Technologies}} \bibinfo{volume}{2020}, \bibinfo{number}{2}
  (\bibinfo{year}{2020}), \bibinfo{pages}{481--498}.
\newblock


\bibitem[\protect\citeauthoryear{Maier and Harr}{Maier and Harr}{2020}]%
        {maier2020dark}
\bibfield{author}{\bibinfo{person}{Maximilian Maier} {and}
  \bibinfo{person}{Rikard Harr}.} \bibinfo{year}{2020}\natexlab{}.
\newblock \showarticletitle{Dark Design Patterns: An End-User Perspective.}
\newblock \bibinfo{journal}{\emph{Human Technology}} \bibinfo{volume}{16},
  \bibinfo{number}{2} (\bibinfo{year}{2020}), \bibinfo{pages}{170--199}.
\newblock


\bibitem[\protect\citeauthoryear{Mathur, Acar, Friedman, Lucherini, Mayer,
  Chetty, and Narayanan}{Mathur et~al\mbox{.}}{2019}]%
        {mathur2019cscw}
\bibfield{author}{\bibinfo{person}{Arunesh Mathur}, \bibinfo{person}{Gunes
  Acar}, \bibinfo{person}{Michael~J. Friedman}, \bibinfo{person}{Elena
  Lucherini}, \bibinfo{person}{Jonathan Mayer}, \bibinfo{person}{Marshini
  Chetty}, {and} \bibinfo{person}{Arvind Narayanan}.}
  \bibinfo{year}{2019}\natexlab{}.
\newblock \showarticletitle{Dark Patterns at Scale: Findings from a Crawl of
  11K Shopping Websites}.
\newblock \bibinfo{journal}{\emph{Proc. ACM Hum.-Comput. Interact.}}
  \bibinfo{volume}{3}, \bibinfo{number}{CSCW}, Article \bibinfo{articleno}{81}
  (\bibinfo{date}{Nov.} \bibinfo{year}{2019}), \bibinfo{numpages}{32}~pages.
\newblock
\showISSN{2573-0142}
\urldef\tempurl%
\url{https://doi.org/10.1145/3359183}
\showDOI{\tempurl}


\bibitem[\protect\citeauthoryear{Mathur, Narayanan, and Chetty}{Mathur
  et~al\mbox{.}}{2018}]%
        {Mathuram}
\bibfield{author}{\bibinfo{person}{Arunesh Mathur}, \bibinfo{person}{Arvind
  Narayanan}, {and} \bibinfo{person}{Marshini Chetty}.}
  \bibinfo{year}{2018}\natexlab{}.
\newblock \showarticletitle{Endorsements on Social Media: An Empirical Study of
  Affiliate Marketing Disclosures on YouTube and Pinterest}.
\newblock \bibinfo{journal}{\emph{Proc. ACM Hum.-Comput. Interact.}}
  \bibinfo{volume}{2}, \bibinfo{number}{CSCW}, Article \bibinfo{articleno}{119}
  (\bibinfo{date}{Nov.} \bibinfo{year}{2018}), \bibinfo{numpages}{26}~pages.
\newblock
\showISSN{2573-0142}
\urldef\tempurl%
\url{https://doi.org/10.1145/3274388}
\showDOI{\tempurl}


\bibitem[\protect\citeauthoryear{Narayanan, Mathur, Chetty, and
  Kshirsagar}{Narayanan et~al\mbox{.}}{2020}]%
        {narayanan2020dark}
\bibfield{author}{\bibinfo{person}{Arvind Narayanan}, \bibinfo{person}{Arunesh
  Mathur}, \bibinfo{person}{Marshini Chetty}, {and} \bibinfo{person}{Mihir
  Kshirsagar}.} \bibinfo{year}{2020}\natexlab{}.
\newblock \showarticletitle{Dark Patterns: Past, Present, and Future}.
\newblock \bibinfo{journal}{\emph{Queue}} \bibinfo{volume}{18},
  \bibinfo{number}{2} (\bibinfo{year}{2020}), \bibinfo{pages}{67--92}.
\newblock


\bibitem[\protect\citeauthoryear{Noggle}{Noggle}{1996}]%
        {noggle1996manipulative}
\bibfield{author}{\bibinfo{person}{Robert Noggle}.}
  \bibinfo{year}{1996}\natexlab{}.
\newblock \showarticletitle{Manipulative actions: a conceptual and moral
  analysis}.
\newblock \bibinfo{journal}{\emph{American Philosophical Quarterly}}
  \bibinfo{volume}{33}, \bibinfo{number}{1} (\bibinfo{year}{1996}),
  \bibinfo{pages}{43--55}.
\newblock


\bibitem[\protect\citeauthoryear{Nouwens, Liccardi, Veale, Karger, and
  Kagal}{Nouwens et~al\mbox{.}}{2020}]%
        {nouwens}
\bibfield{author}{\bibinfo{person}{Midas Nouwens}, \bibinfo{person}{Ilaria
  Liccardi}, \bibinfo{person}{Michael Veale}, \bibinfo{person}{David Karger},
  {and} \bibinfo{person}{Lalana Kagal}.} \bibinfo{year}{2020}\natexlab{}.
\newblock \showarticletitle{Dark Patterns after the GDPR: Scraping Consent
  Pop-Ups and Demonstrating Their Influence}. In
  \bibinfo{booktitle}{\emph{Proceedings of the 2020 CHI Conference on Human
  Factors in Computing Systems}} (Honolulu, HI, USA)
  \emph{(\bibinfo{series}{CHI '20})}. \bibinfo{publisher}{Association for
  Computing Machinery}, \bibinfo{address}{New York, NY, USA},
  \bibinfo{pages}{1–13}.
\newblock
\showISBNx{9781450367080}
\urldef\tempurl%
\url{https://doi.org/10.1145/3313831.3376321}
\showDOI{\tempurl}


\bibitem[\protect\citeauthoryear{of~State}{of~State}{2020}]%
        {caballot}
\bibfield{author}{\bibinfo{person}{California~Secretary of State}.}
  \bibinfo{year}{2020}\natexlab{}.
\newblock \bibinfo{title}{Qualified Statewide Ballot Measures}.
\newblock
  \bibinfo{howpublished}{\url{https://www.oag.ca.gov/system/files/initiatives/pdfs/19-0021A1\%20\%28Consumer\%20Privacy\%20-\%20Version\%203\%29_1.pdf}}.
\newblock
\newblock
\shownote{Accessed September 9, 2020.}


\bibitem[\protect\citeauthoryear{Olson and Kellogg}{Olson and Kellogg}{2014}]%
        {olson2014ways}
\bibfield{author}{\bibinfo{person}{Judith~S. Olson} {and}
  \bibinfo{person}{Wendy~A. Kellogg}.} \bibinfo{year}{2014}\natexlab{}.
\newblock \bibinfo{booktitle}{\emph{Ways of Knowing in HCI}}.
\newblock \bibinfo{publisher}{Springer Publishing Company, Incorporated},
  \bibinfo{address}{USA}.
\newblock
\showISBNx{1493903772}


\bibitem[\protect\citeauthoryear{on~Informatics and (CNIL)}{on~Informatics and
  (CNIL)}{2020}]%
        {cnil}
\bibfield{author}{\bibinfo{person}{National~Commission on Informatics} {and}
  \bibinfo{person}{Liberty (CNIL)}.} \bibinfo{year}{2020}\natexlab{}.
\newblock \bibinfo{title}{Shaping Choices in the Digital World}.
\newblock
  \bibinfo{howpublished}{\url{https://linc.cnil.fr/sites/default/files/atoms/files/cnil_ip_report_06_shaping_choices_in_the_digital_world.pdf}}.
\newblock
\newblock
\shownote{Accessed September 9, 2020.}


\bibitem[\protect\citeauthoryear{Pitofsky}{Pitofsky}{1977}]%
        {pitofsky}
\bibfield{author}{\bibinfo{person}{Robert Pitofsky}.}
  \bibinfo{year}{1977}\natexlab{}.
\newblock \showarticletitle{Beyond Nader: Consumer Protection and the
  Regulation of Advertising}.
\newblock \bibinfo{journal}{\emph{Harvard Law Review}} \bibinfo{volume}{90},
  \bibinfo{number}{4} (\bibinfo{year}{1977}), \bibinfo{pages}{661--701}.
\newblock
\showISSN{0017811X}
\urldef\tempurl%
\url{http://www.jstor.org/stable/1340281}
\showURL{%
\tempurl}


\bibitem[\protect\citeauthoryear{Rudinow}{Rudinow}{1978}]%
        {Rudinow1978}
\bibfield{author}{\bibinfo{person}{Joel Rudinow}.}
  \bibinfo{year}{1978}\natexlab{}.
\newblock \showarticletitle{Manipulation}.
\newblock \bibinfo{journal}{\emph{Ethics}} \bibinfo{volume}{88},
  \bibinfo{number}{4} (\bibinfo{year}{1978}), \bibinfo{pages}{338--347}.
\newblock
\urldef\tempurl%
\url{https://doi.org/10.1086/292086}
\showDOI{\tempurl}


\bibitem[\protect\citeauthoryear{Sch{\"u}ll}{Sch{\"u}ll}{2014}]%
        {schull2014addiction}
\bibfield{author}{\bibinfo{person}{Natasha~Dow Sch{\"u}ll}.}
  \bibinfo{year}{2014}\natexlab{}.
\newblock \bibinfo{booktitle}{\emph{Addiction by design: Machine gambling in
  Las Vegas}}.
\newblock \bibinfo{publisher}{Princeton University Press}.
\newblock


\bibitem[\protect\citeauthoryear{Shaw}{Shaw}{2019}]%
        {wise-nudge}
\bibfield{author}{\bibinfo{person}{Simon Shaw}.}
  \bibinfo{year}{2019}\natexlab{}.
\newblock \bibinfo{title}{Consumers Are Becoming Wise to Your Nudge}.
\newblock
  \bibinfo{howpublished}{\url{https://behavioralscientist.org/consumers-are-becoming-wise-to-your-nudge/}}.
\newblock
\newblock
\shownote{Accessed September 9, 2020.}


\bibitem[\protect\citeauthoryear{Soe, Nordberg, Guribye, and Slavkovik}{Soe
  et~al\mbox{.}}{2020}]%
        {circumvention}
\bibfield{author}{\bibinfo{person}{Than~Htut Soe}, \bibinfo{person}{Oda~Elise
  Nordberg}, \bibinfo{person}{Frode Guribye}, {and} \bibinfo{person}{Marija
  Slavkovik}.} \bibinfo{year}{2020}\natexlab{}.
\newblock \showarticletitle{Circumvention by Design -- Dark Patterns in Cookie
  Consent for Online News Outlets}. In \bibinfo{booktitle}{\emph{Proceedings of
  the 11th Nordic Conference on Human-Computer Interaction: Shaping
  Experiences, Shaping Society}} (Tallinn, Estonia)
  \emph{(\bibinfo{series}{NordiCHI '20})}. \bibinfo{publisher}{Association for
  Computing Machinery}, \bibinfo{address}{New York, NY, USA}, Article
  \bibinfo{articleno}{19}, \bibinfo{numpages}{12}~pages.
\newblock
\showISBNx{9781450375795}
\urldef\tempurl%
\url{https://doi.org/10.1145/3419249.3420132}
\showDOI{\tempurl}


\bibitem[\protect\citeauthoryear{Solove}{Solove}{2005}]%
        {solove2005taxonomy}
\bibfield{author}{\bibinfo{person}{Daniel~J Solove}.}
  \bibinfo{year}{2005}\natexlab{}.
\newblock \showarticletitle{A taxonomy of privacy}.
\newblock \bibinfo{journal}{\emph{U. Pa. L. Rev.}}  \bibinfo{volume}{154}
  (\bibinfo{year}{2005}), \bibinfo{pages}{477}.
\newblock


\bibitem[\protect\citeauthoryear{Sunstein}{Sunstein}{2017}]%
        {sunstein_2017}
\bibfield{author}{\bibinfo{person}{Cass~R. Sunstein}.}
  \bibinfo{year}{2017}\natexlab{}.
\newblock \showarticletitle{Nudges that fail}.
\newblock \bibinfo{journal}{\emph{Behavioural Public Policy}}
  \bibinfo{volume}{1}, \bibinfo{number}{1} (\bibinfo{year}{2017}),
  \bibinfo{pages}{4–25}.
\newblock
\urldef\tempurl%
\url{https://doi.org/10.1017/bpp.2016.3}
\showDOI{\tempurl}


\bibitem[\protect\citeauthoryear{Sunstein}{Sunstein}{2020}]%
        {sunstein2020sludge}
\bibfield{author}{\bibinfo{person}{Cass~R. Sunstein}.}
  \bibinfo{year}{2020}\natexlab{}.
\newblock \showarticletitle{Sludge Audits}.
\newblock \bibinfo{journal}{\emph{Behavioural Public Policy}}
  (\bibinfo{year}{2020}), \bibinfo{pages}{1–20}.
\newblock
\urldef\tempurl%
\url{https://doi.org/10.1017/bpp.2019.32}
\showDOI{\tempurl}


\bibitem[\protect\citeauthoryear{Susser, Roessler, and Nissenbaum}{Susser
  et~al\mbox{.}}{2020}]%
        {susser-online-manipulation}
\bibfield{author}{\bibinfo{person}{Daniel Susser}, \bibinfo{person}{Beate
  Roessler}, {and} \bibinfo{person}{Helen Nissenbaum}.}
  \bibinfo{year}{2020}\natexlab{}.
\newblock \showarticletitle{Online Manipulation: Hidden Influences in a Digital
  World}.
\newblock \bibinfo{journal}{\emph{Georgetown Law Technology Review}}
  \bibinfo{volume}{4} (\bibinfo{year}{2020}), 45.
\newblock


\bibitem[\protect\citeauthoryear{Sweller}{Sweller}{2018}]%
        {sweller2018measuring}
\bibfield{author}{\bibinfo{person}{John Sweller}.}
  \bibinfo{year}{2018}\natexlab{}.
\newblock \showarticletitle{Measuring cognitive load}.
\newblock \bibinfo{journal}{\emph{Perspectives on medical education}}
  \bibinfo{volume}{7}, \bibinfo{number}{1} (\bibinfo{year}{2018}),
  \bibinfo{pages}{1--2}.
\newblock


\bibitem[\protect\citeauthoryear{Thaler}{Thaler}{2018}]%
        {Thaler431}
\bibfield{author}{\bibinfo{person}{Richard~H. Thaler}.}
  \bibinfo{year}{2018}\natexlab{}.
\newblock \showarticletitle{Nudge, not sludge}.
\newblock \bibinfo{journal}{\emph{Science}} \bibinfo{volume}{361},
  \bibinfo{number}{6401} (\bibinfo{year}{2018}), \bibinfo{pages}{431--431}.
\newblock
\showISSN{0036-8075}
\urldef\tempurl%
\url{https://science.sciencemag.org/content/361/6401/431.full.pdf}
\showURL{%
\tempurl}


\bibitem[\protect\citeauthoryear{Thaler and Sunstein}{Thaler and
  Sunstein}{2009}]%
        {thaler2009nudge}
\bibfield{author}{\bibinfo{person}{Richard~H Thaler} {and}
  \bibinfo{person}{Cass~R Sunstein}.} \bibinfo{year}{2009}\natexlab{}.
\newblock \bibinfo{booktitle}{\emph{Nudge: Improving decisions about health,
  wealth, and happiness}}.
\newblock \bibinfo{publisher}{Penguin}.
\newblock


\bibitem[\protect\citeauthoryear{Tversky and Kahneman}{Tversky and
  Kahneman}{1979}]%
        {tversky1979prospect}
\bibfield{author}{\bibinfo{person}{Amos Tversky} {and} \bibinfo{person}{Daniel
  Kahneman}.} \bibinfo{year}{1979}\natexlab{}.
\newblock \showarticletitle{Prospect theory: An analysis of decision under
  risk}.
\newblock \bibinfo{journal}{\emph{Econometrica}} \bibinfo{volume}{47},
  \bibinfo{number}{2} (\bibinfo{year}{1979}), \bibinfo{pages}{263--291}.
\newblock


\bibitem[\protect\citeauthoryear{Utz, Degeling, Fahl, Schaub, and Holz}{Utz
  et~al\mbox{.}}{2019}]%
        {utz2019informed}
\bibfield{author}{\bibinfo{person}{Christine Utz}, \bibinfo{person}{Martin
  Degeling}, \bibinfo{person}{Sascha Fahl}, \bibinfo{person}{Florian Schaub},
  {and} \bibinfo{person}{Thorsten Holz}.} \bibinfo{year}{2019}\natexlab{}.
\newblock \showarticletitle{(Un)Informed Consent: Studying GDPR Consent Notices
  in the Field}. In \bibinfo{booktitle}{\emph{Proceedings of the 2019 ACM
  SIGSAC Conference on Computer and Communications Security}} (London, United
  Kingdom) \emph{(\bibinfo{series}{CCS '19})}. \bibinfo{publisher}{Association
  for Computing Machinery}, \bibinfo{address}{New York, NY, USA},
  \bibinfo{pages}{973–990}.
\newblock
\showISBNx{9781450367479}
\urldef\tempurl%
\url{https://doi.org/10.1145/3319535.3354212}
\showDOI{\tempurl}


\bibitem[\protect\citeauthoryear{Waldman}{Waldman}{2020}]%
        {WALDMAN2020105}
\bibfield{author}{\bibinfo{person}{Ari~Ezra Waldman}.}
  \bibinfo{year}{2020}\natexlab{}.
\newblock \showarticletitle{Cognitive biases, dark patterns, and the ‘privacy
  paradox’}.
\newblock \bibinfo{journal}{\emph{Current Opinion in Psychology}}
  \bibinfo{volume}{31} (\bibinfo{year}{2020}), \bibinfo{pages}{105--109}.
\newblock
\showISSN{2352-250X}
\urldef\tempurl%
\url{https://doi.org/10.1016/j.copsyc.2019.08.025}
\showDOI{\tempurl}
\newblock
\shownote{Privacy and Disclosure, Online and in Social Interactions.}


\bibitem[\protect\citeauthoryear{Warner and Fischer}{Warner and
  Fischer}{2019}]%
        {detour-act}
\bibfield{author}{\bibinfo{person}{Mark Warner} {and} \bibinfo{person}{Debra
  Fischer}.} \bibinfo{year}{2019}\natexlab{}.
\newblock \bibinfo{title}{Senators Introduce Bipartisan Legislation to Ban
  Manipulative ``Dark Patterns''}.
\newblock
  \bibinfo{howpublished}{\url{https://www.fischer.senate.gov/public/index.cfm/2019/4/senators-introduce-bipartisan-legislation-to-ban-manipulative-dark-patterns}}.
\newblock
\newblock
\shownote{Accessed September 9, 2020.}


\bibitem[\protect\citeauthoryear{Westin and Chiasson}{Westin and
  Chiasson}{2019}]%
        {fomo}
\bibfield{author}{\bibinfo{person}{Fiona Westin} {and} \bibinfo{person}{Sonia
  Chiasson}.} \bibinfo{year}{2019}\natexlab{}.
\newblock \showarticletitle{Opt out of Privacy or ``Go Home'': Understanding
  Reluctant Privacy Behaviours through the FoMO-Centric Design Paradigm}. In
  \bibinfo{booktitle}{\emph{Proceedings of the New Security Paradigms
  Workshop}} (San Carlos, Costa Rica) \emph{(\bibinfo{series}{NSPW '19})}.
  \bibinfo{publisher}{Association for Computing Machinery},
  \bibinfo{address}{New York, NY, USA}, \bibinfo{pages}{57–67}.
\newblock
\showISBNx{9781450376471}
\urldef\tempurl%
\url{https://doi.org/10.1145/3368860.3368865}
\showDOI{\tempurl}


\bibitem[\protect\citeauthoryear{Wilkinson}{Wilkinson}{2013}]%
        {wilkinson2013nudging}
\bibfield{author}{\bibinfo{person}{T.~Martin Wilkinson}.}
  \bibinfo{year}{2013}\natexlab{}.
\newblock \showarticletitle{Nudging and manipulation}.
\newblock \bibinfo{journal}{\emph{Political Studies}} \bibinfo{volume}{61},
  \bibinfo{number}{2} (\bibinfo{year}{2013}), \bibinfo{pages}{341--355}.
\newblock


\bibitem[\protect\citeauthoryear{Wood}{Wood}{2014}]%
        {wood2014coercion}
\bibfield{author}{\bibinfo{person}{Allen Wood}.}
  \bibinfo{year}{2014}\natexlab{}.
\newblock \showarticletitle{Coercion, manipulation, exploitation}.
\newblock \bibinfo{journal}{\emph{Manipulation: Theory and practice}}
  \bibinfo{volume}{1} (\bibinfo{year}{2014}), \bibinfo{pages}{17--50}.
\newblock


\bibitem[\protect\citeauthoryear{Zagal, Bj{\"o}rk, and Lewis}{Zagal
  et~al\mbox{.}}{2013}]%
        {zagal2013dark}
\bibfield{author}{\bibinfo{person}{Jos{\'e}~P Zagal}, \bibinfo{person}{Staffan
  Bj{\"o}rk}, {and} \bibinfo{person}{Chris Lewis}.}
  \bibinfo{year}{2013}\natexlab{}.
\newblock \showarticletitle{Dark patterns in the design of games}. In
  \bibinfo{booktitle}{\emph{Foundations of Digital Games 2013}}.
  \bibinfo{publisher}{Society for the Advancement of the Science of Digital
  Games}, \bibinfo{address}{Santa Cruz, CA}, 8.
\newblock


\end{thebibliography}
